\documentclass[12pt]{article}
\usepackage{cmap}
\usepackage[utf8]{inputenc}
\usepackage[T2A]{fontenc}
\usepackage[english]{babel}
\inputencoding{utf8}
\usepackage{amsmath,amssymb,amsthm,graphicx}
\graphicspath{{..},{pic}}

\usepackage[pdfpagemode=UseOutlines,
bookmarks=true, bookmarksopenlevel=3, bookmarksopen=true,
bookmarksnumbered=true, unicode=true,
pdffitwindow=true,pdfpagelayout=SinglePage,pdfhighlight=/P,
colorlinks=true,linkcolor=blue,citecolor=blue,urlcolor=blue]{hyperref}

\usepackage[left=2cm, right=2cm, top=2cm, bottom=2cm]{geometry}

\usepackage[width=\textwidth,indention=0mm,justification=centerlast,
 font=small,labelfont=bf,labelsep=period]{caption}

\numberwithin{equation}{section}

\theoremstyle{plain}
\newtheorem{proposition}{Proposition}
\theoremstyle{remark}
\newtheorem{remark}{Remark}
\newtheorem{example}{Example}

\title{Quasi-stationary solutions of the dP$_\text{I}$ equation}

\author{V.E.\:Adler\thanks{L.D.\:Landau Institute for Theoretical Physics, Chernogolovka, Russian Federation.},\quad 
B.I.\:Suleimanov\thanks{Institute of Mathematics, Ufa Federal Research Centre, Russian Academy of Sciences, 112, Chernyshevsky Street, Ufa 450008, Russian Federation.}} 

\date{November 20, 2026}

\begin{document}
\maketitle

\begin{abstract} 
We study real quasi-stationary solutions of the discrete Painlev\'e-I equation (dP$_\text{I}$) which defines a reduction of the Volterra lattice and the B\"acklund transformations for the Painlev\'e-IV equation (P$_\text{IV}$). It is shown that the quasi-stationary asymptotics for large $|n|$ is preserved under the evolution of the solution with respect to the parameter $t$ governed by P$_\text{IV}$ and, conversely, the quasi-stationary asymptotics for large $|t|$ is preserved under the evolution with respect to $n$ governed by dP$_\text{I}$. This makes it possible to speak of a common quasi-stationary surface and, in the case of uniform asymptotics, of common quasi-stationary solutions for these equations. Examples (exact and numerical) are given of solutions demonstrating quasi-stationary and pole behavior in different regions of the $(n,t)$-plane. The existence of global quasi-stationary solutions is shown numerically. A uniformly valid self-similar asymptotics is constructed in the transition zone between the branches of the quasi-stationary curve, related to Hastings--McLeod type solutions of the Painlev\'e-II equation (P$_\text{II}$).
\medskip

\end{abstract}

\section{Introduction}

The discrete Painlev\'e-I equation (dP$_\text{I}$)
\begin{equation}\label{dP1}
 u_n(u_{n-1}+u_n+u_{n+1})=2tu_n+\beta_n,\qquad \beta_n=n-b-(-1)^nc
\end{equation}
arises in the study of polynomials $P_n(x)$, $n\ge0$, orthogonal on the real line with respect to the weight $|x|^{-2b}e^{-x^4+2tx^2}$. For the first time this equation appeared in the paper by Shohat (1939) \cite{Shohat_1939}, at $b=c=t=0$. The generalization for $\beta_n=n-(1-(-1)^n)b$ was introduced by Freud (1976) \cite{Freud_1976}, the parameter $t$ was introduced by Bonan \& Nevai (1984) \cite{Bonan_Nevai_1984}. In this context, the equation is considered only for $n\geqslant 0$ (of course, up to a shift of the numbering), the variables $u_n$ play the role of coefficients in the three-term recurrence relation $P_{n+1}=xP_n-\frac{1}{2}u_nP_{n-1}$ and must satisfy the conditions $u_n>0$ for $n>0$ and $u_0=0$, which implies that $c=-b$. It turns out that there exists exactly one solution of dP$_\text{I}$ with these properties. For $t=0$, this was proved by Lew \& Quarles (1983) \cite{Lew_Quarles_1983} (even for more general $\beta_n$); for $b=0$, it was proved in \cite{Bonan_Nevai_1984}; and for arbitrary $b<1/2$ and $t$, it was proved in \cite{Clarkson_Jordaan_Kelil_2015, Clarkson_Jordaan_2018}. Equations with more general quadratic terms were studied in \cite{Alsulami_Nevai_Szabados_Assche_2015}. In the early 90s, dP$_\text{I}$ was rediscovered in the framework of the theory of integrable equations and matrix models \cite{Brezin_Kazakov_1990, Its_Kitaev_Fokas_1990, Fokas_Its_Kitaev_1991, Fokas_Its_Kitaev_1992}, see also the paper \cite{Magnus_1995}, which discusses the connection to the theory of orthogonal polynomials, including cases with more general weight functions, and provides an extensive bibliography on the subject. In these same papers, the connection with the Painlev\'e-IV equation (P$_\text{IV}$) and with the isomonodromic deformation method was established. In this context, the dP$_\text{I}$ equation can be considered for all $n$, which removes the restriction $c=-b$ and the positivity condition for $u_n$. The three-term recurrence relation for orthogonal polynomials is interpreted as one of the auxiliary linear problems, where $x$ plays the role of the spectral parameter. 

Let us present here the key formulas from the cited papers, which are needed in what follows. All variables and parameters are assumed to be real. The equation (\ref{dP1}) defines a constraint compatible with the Volterra lattice
\begin{equation}\label{VL}
 u'_n=u_n(u_{n+1}-u_{n-1}),
\end{equation}
where the prime denotes differentiation with respect to the parameter $t$. This is checked by a direct computation: let $A_n=u_n(u_{n-1}+u_n+u_{n+1})-2tu_n-\beta_n$, then the differentiation in virtue of (\ref{VL}) gives $A'_n=u_n(A_{n+1}-A_{n-1})$, whence it follows that if all $A_n=0$, then $A'_n=0$ as well, which means that the constraint is preserved. 

\begin{remark} 
The derivations
\begin{gather*}
 \partial_{t_2}(u_n)=u_n(u_{n+1}(u_{n+2}+u_{n+1}+u_n)-u_{n-1}(u_n+u_{n-1}+u_{n-2})),\\
 \partial_{\tau}(u_n)=tu_n(u_{n+1}-u_{n-1})+u_n
\end{gather*}
commute with (\ref{VL}): $[\partial_t,\partial_{t_2}]=[\partial_t,\partial_{\tau}]=0$, that is, they define symmetries of this lattice---a higher symmetry and a classical one corresponding to the scaling transformation. The stationary equation of any symmetry is an invariant manifold, and the equation (\ref{dP1}) arises exactly in this way from the equation $\partial_{t_2}(u_n)-2\partial_{\tau}(u_n)=0$ (the simplest example of the so-called discrete string equation). The parameters $b$ and $c$ appear as constants of integration under the reduction of order after dividing the equation by $u_n$; the fact that they must not depend on $t$ is checked directly. Replacing $\partial_{t_2}$ in this construction by higher order symmetries (that is, depending on a larger number of variables $u_{n\pm k}$), one obtains an infinite hierarchy of higher-order analogs of the dP$_\text{I}$ equation, see, e.g. \cite{Cresswell_Joshi_1999}. 
\end{remark} 

By virtue of equations (\ref{dP1}) and (\ref{VL}), the pair $(u_n,u_{n+1})$ for any $n$ satisfies the system of ODEs
\begin{equation}\label{P4-sys}
 \left\{\begin{aligned}
  u'_n &= u_n(2u_{n+1}+u_n)-2tu_n-\beta_n,\\
  u'_{n+1} &= -u_{n+1}(u_{n+1}+2u_n)+2tu_{n+1}+\beta_{n+1}.
 \end{aligned}\right.  
\end{equation}
An alternative proof of the compatibility of (\ref{dP1}) and (\ref{VL}) consists in checking that the form of this system is preserved under the maps $(u_{n-1},u_n)\leftrightarrow(u_n,u_{n+1})$ defined by solving (\ref{dP1}) with respect to $u_{n-1}$ or $u_{n+1}$. Eliminating either of the variables $u_n$ or $u_{n+1}$ reduces (\ref{P4-sys}) to the P$_\text{IV}$ equation for the other variable (note that the canonical form of P$_\text{IV}$ differs in the sign of $t$ and in the notation for the coefficients):
\begin{equation}\label{P4}
 u''=\frac{(u')^2}{2u}+\frac{3}{2}u^3-4tu^2+(2t^2+\alpha)u-\frac{\beta^2}{2u}
\end{equation}
with the parameter values in the equation for $u_n$ given by
\begin{equation}\label{alphabet}
 \beta_n=n-b-(-1)^nc,\qquad \alpha_n=2\beta_{n+1}-\beta_n-2=n-b+3(-1)^nc.
\end{equation}
The system (\ref{P4-sys}) is equivalent to the mutually inverse B\"acklund transformations $B$ and $B^{-1}$ acting on solutions and parameters of P$_\text{IV}$ according to the formulas 
\begin{equation}\label{BT-gen}
 B^{\pm1}:~ \alpha_{n\pm1}=\frac{3\beta_n-\alpha_n}{2}\pm1,\quad 
 \beta_{n\pm1}=\frac{\alpha_n+\beta_n}{2}\pm1,\quad 
 u_{n\pm1}= \frac{\pm u'_n-u^2_n+2tu_n+\beta_n}{2u_n}.
\end{equation}
Thus, the evolution of solutions of dP$_\text{I}$ with respect to the parameter $t$ is described by the P$_\text{IV}$ equation and, conversely, dP$_\text{I}$ defines a sequence of transformations acting on solutions of P$_\text{IV}$. 

\begin{remark}\label{rem:BT} 
The B\"acklund transformations for P$_\text{IV}$ were discovered by Lukashevich (1967) \cite{Lukashevich_1967}, see also the book \cite{Gromak_Laine_Shimomura_2002}, which contains a detailed bibliography. We will use only the transformations (\ref{BT-gen}), but it is worth noting that the full group of B\"acklund transformations for the P$_\text{IV}$ equation is larger. This is related to the fact that the parameter $\beta$ in (\ref{P4}) is defined up to a sign. As the generators of the group one can take the transformations 
\[
 R:(\alpha,\beta,u)\mapsto(\alpha,-\beta,u),\qquad
 S:(\alpha,\beta,u)\mapsto\left(\frac{3\beta-\alpha}{2}+1, 
  -\frac{\alpha+\beta}{2}-1, \frac{u'-u^2+2tu+\beta}{2u}\right),
\]
satisfying the identities
\[
 R^2=S^3=(SRSRS)^3=\operatorname{id},
\]
and for generic solutions of P$_\text{IV}$ there are no other independent identities. Solutions generated by the transformation $S$ satisfy the 3-periodic dressing chain \cite{Veselov_Shabat_1993, Adler_1994}. Although the transformation $R$ is trivial by itself, by means of it a two-dimensional lattice of P$_\text{IV}$ solutions is generated (more precisely, the kagome lattice, see, e.g. \cite{Bassom_Clarkson_Hicks_1995}), whereas the transformations $B=RS$ and $B^{-1}=S^2R$ generate only a one-dimensional sequence of solutions.
\end{remark} 

As previously noted, from the point of view of the theory of orthogonal polynomials, of interest are not all solutions of dP$_\text{I}$, but only those satisfying the termination condition $u_0=0$, which reduces the system (\ref{P4-sys}) to the Riccati equation for $u_1$ and, upon linearization, to the Hermite equation. The B\"acklund transformations generate families of special solutions of P$_\text{IV}$, which were studied in detail in a number of papers; we mention only \cite{Gromak_1978, Bassom_Clarkson_Hicks_McLeod_1992, Bassom_Clarkson_Hicks_1993, Bassom_Clarkson_Hicks_1995}. Thus, the corresponding common solutions of dP$_\text{I}$ and P$_\text{IV}$ are expressed in terms of classical special functions of the parameter $t$, which makes it possible, eventually, to effectively single out the unique solution satisfying the positivity property, by deriving an exact formula for the variable $u_1$. This solution is also characterized by its asymptotic behavior $u_n\sim\sqrt{n/3}$ as $n\to\infty$, which was shown already in \cite{Lew_Quarles_1983}, see also the paper \cite{Nevai_1984}, where the asymptotics of the orthogonal polynomials themselves was studied, and \cite{Bleher_Its_1999, Bleher_Its_2003}, where the asymptotics of solutions was studied in the framework of the Riemann--Hilbert approach.

In this paper we study generalizations of this special dP$_\text{I}$--P$_\text{IV}$ solution obtained by dropping the zero termination condition. The starting point of our approach is the condition of square-root asymptotics as $n\to\pm\infty$. We present the results of numerical experiments, which show that such solutions exist, at least, for $t>0$ and small values of $c$. Moreover, the positivity property holds in the following form: the solutions are positive for $n>0$ and have alternating signs for $n\leqslant0$ (it is easily seen from the equation (\ref{dP1}) itself that the solution cannot be of a fixed sign as $n\to-\infty$). A rigorous proof of the existence of such solutions should be more difficult than in the case of zero termination, due to their transcendence, and remains an open problem. In addition to the numerical computation, we also carry out a rather detailed study of formal asymptotic expansions in $n$ and in $t$, and check their consistency.

Let us briefly describe the content of the paper. In Section \ref{s:QSC}, the quasi-stationary curves for the dP$_\text{I}$ and P$_\text{IV}$ equations are defined as the loci of equilibrium points for the corresponding autonomous equation with frozen independent variable ($n$ or $t$). A similar idea was considered in \cite{Ercolani_Lega_Tippings_2022}. Quasi-stationary solutions are defined as asymptotically close to these curves, which leads to the aforementioned square-root asymptotics. An important circumstance is that the compatibility of the equations dP$_\text{I}$ and P$_\text{IV}$ makes it possible to define also their common quasi-stationary surface. The attractive idea is that the evolution governed by one of these equations should preserve the quasi-stationarity for the solutions of the other one. We verify this property at the level of formal asymptotics. In Section \ref{s:t-series}, asymptotic expansions for quasi-stationary solutions of P$_\text{IV}$ are analyzed and it is shown that their structure is preserved under the shift in $n$ due to dP$_\text{I}$; conversely, in Section \ref{s:n-series} it is shown that the asymptotic expansions for dP$_\text{I}$ are preserved under the evolution in $t$ due to P$_\text{IV}$. Thus, the results of these sections support the possibility of existence of a common quasi-stationary solution of dP$_\text{I}$ and P$_\text{IV}$. Section \ref{s:bvp} describes the boundary value problems and difference schemes for constructing quasi-stationary solutions. 

The quasi-stationarity property can also be interpreted in a broader sense, as the closeness to the quasi-stationary curve on some finite interval. There exist, of course, a lot of such solutions; some simple examples are given in Section \ref{s:examples}. Section \ref{s:zero} is devoted to the case of zero termination; we present the known results for $n>0$ and, possibly, new results for $n<0$. In Section \ref{s:regular}, we return to our primary objective: the construction of globally regular solutions. The necessary conditions for their existence are derived through an analysis of the asymptotic series from Section \ref{s:t-series}; furthermore, we present results from numerical experiments that support the conjecture that these conditions are also sufficient.

Section \ref{s:self} contains one more type of asymptotic series, which takes into account the self-similar structure of the solutions, for the values $c<1/2$; this makes it possible to study their behavior in the transition region between the branches of the quasi-stationary curves. The leading terms of such self-similar asymptotics are described by special Hastings--McLeod type solutions of the Painlev\'e-II equation (P$_\text{II}$). This gives a certain generalization of the results by Bleher \& Its (2003) \cite{Bleher_Its_2003}.

\section{Quasi-stationary curves and surfaces}\label{s:QSC}

\subsection{Basic definitions}

\begin{figure}[t!]
\centerline{\includegraphics[width=0.8\textwidth]{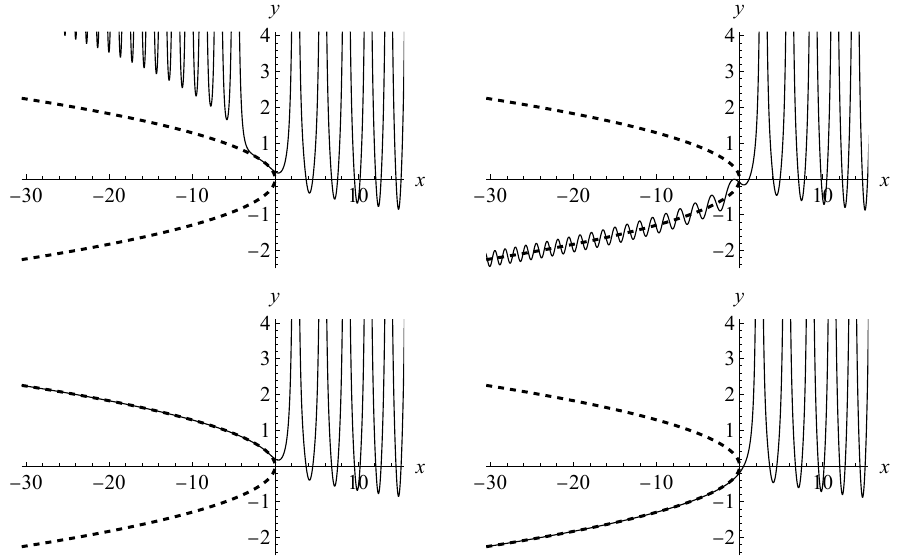}}
\caption{Solutions of P$_\text{I}$.}
\label{fig:P1-solutions} 
\end{figure}

Consider a non-autonomous ODE with slowly varying coefficients. Its quasi-stationary curve is defined as the locus of the equilibrium points of the autonomous equation with the frozen independent variable. For example, for the P$_\text{I}$ equation
\begin{equation}\label{P1}
 y''=6y^2+x   
\end{equation}
one should pass to the autonomous equation $d^2y/d\xi^2=6y^2+x$, which depends on $x$ as a parameter; the equilibrium points of this equation form the parabola $6y^2+x=0$, which serves as the quasi-stationary curve. Generic solutions (see Fig.~\ref{fig:P1-solutions}, top) are far from this parabola. For $x>0$, the quasi-stationary curve is not defined, and in this region all solutions develop poles. For $x<0$, the lower branch of the parabola, $y=-\sqrt{-x/6}$, is stable, since the linearized equation $Y''=12yY$ possesses purely imaginary eigenvalues along this branch. Consequently, for a certain range of initial data, the solutions exhibit an oscillatory behavior on this background. The upper branch is unstable, and the solutions tend to leave it either into the pole regime or into the regime of oscillations along the lower branch. At the same time, there exist special solutions shown in Fig.~\ref{fig:P1-solutions}, bottom. The solution approaching the lower branch of the parabola without oscillations is the famous tritronqu\'ee solution which is unique. The solutions approaching the upper branch form an unstable one-parameter family. In general, solutions asymptotically approaching one of the branches of the quasi-stationary curve will be called quasi-stationary. As another example, in Section \ref{s:self} we consider solutions to the P$_\text{II}$ equation of the Hastings--McLeod type, for which a quasi-stationary curve determines the leading terms of the asymptotic expansion for large values of the argument.

The definitions for the discrete case are similar. Consider an equation of the form
\[
 P_n(u_{n-1},u_n,u_{n+1})=0,
\]
where the left hand side depends on $n$ explicitly (polynomially, for instance). Assuming that this dependence can be considered slow, in a sense, we replace this equation by an autonomous one, keeping the dependence on $n$ in the coefficients, as on a parameter:
\[
 P_n(u_{k-1},u_k,u_{k+1})=0.
\] 
In this context, the variable $n$ can be regarded as an arbitrary real rather than an integer. The fixed points of the corresponding map (or, possibly, multivalued correspondence)
\[
 T_n:(u_{k-1},u_k)\mapsto(u_k,u_{k+1})
\] 
are defined as the solutions of the equation $\Phi_n(u)=P_n(u,u,u)=0$, which defines the quasi-stationary curve. An important difference from the continuous case is that, for difference equations, it is natural to consider fixed points not only of the map $T_n$ itself, but also of its powers $T^m_n$, that is, $m$-periodic orbits of $T_n$. Here, if $d$ is a divisor of $m$, then the $d$-periodic orbits can be discarded. This defines quasi-stationary curves of higher degree. Solutions asymptotically close to certain branches of these curves will be called quasi-stationary.

\subsection{The \texorpdfstring{dP$_\text{I}$}{dP-I} equation}\label{s:dP1}

This definition is not directly applicable to the general form of dP$_\text{I}$ (\ref{dP1}), due to the oscillating term $(-1)^nc$, which cannot be considered slowly varying. For simplicity, we consider first the equation with $c=0$:
\begin{equation}\label{dP1-0}
 u_n(u_{n-1}+u_n+u_{n+1})-2tu_n-\nu=0,\qquad \nu:=n-b.
\end{equation}
For it, the quasi-stationary curve of the first degree is the parabola
\begin{equation}\label{curve1}
 C_1:~~ 3u^2-2tu-\nu=0
\end{equation}
(this is reminiscent of the example for P$_\text{I}$; however, it should be noted right away that the resemblance to the P$_\text{I}$ equation is quite distant). A quasi-stationary solution must be close, as $n\to+\infty$, either to the upper or to the lower branch of this parabola: it is not difficult to verify that solutions tending to these branches alternately do not exist. Thus, two types of quasi-stationary solutions are possible here, but, since dP$_\text{I}$ is invariant under the change $\tilde u_n=-u_n$, $\tilde t=-t$, in what follows we can restrict ourselves, without loss of generality, to the consideration of solutions tending to the upper branch.

\begin{figure}[t!]
\centerline{\includegraphics[width=0.98\textwidth]{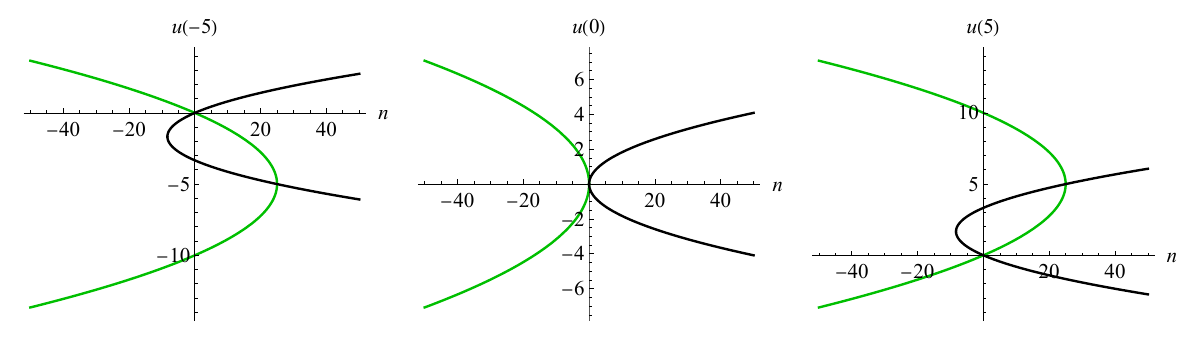}}
\caption{Quasi-stationary curves for dP$_\text{I}$ at $b=c=0$, at different values of $t$\\ (black line: $C_1$, green line: $C_2$).}
\label{fig:dP1-curve12} 
\end{figure}

To find the fixed points of the map $T^2_n$, write out the autonomous equations at two neighboring points, setting $u_{k+2}=u_k=u$ and $u_{k-1}=u_{k+1}=v$:
\[
 u(2v+u)-2tu-\nu=0,\qquad v(2u+v)-2tv-\nu=0.
\]
The elimination of $v$ leads to the equation
\begin{equation}\label{curve12}
 (3u^2-2tu-\nu)(u^2-2tu+\nu)=0.
\end{equation}
The first factor corresponds to the fixed points of the map $T_n$ itself, which we have already found. Equating the second factor to zero, we obtain the parabola
\begin{equation}\label{curve2}
 C_2:~~ u^2-2tu+\nu=0,
\end{equation}
which corresponds to the points of period 2. It is easy to check that, in contrast with the parabola $C_1$, a solution close to $C_2$ as $n\to-\infty$ must be a sequence whose terms tend to the upper and lower branches alternately. Thus, in this case two types of quasi-stationary solutions are also possible, depending on which branch the variables with even numbers converge to. Both curves (\ref{curve1}) and (\ref{curve2}) are shown in Fig.~\ref{fig:dP1-curve12}. Note that they intersect at the points $\nu=0$, $u=0$ and $\nu=t^2$, $u=t$, the latter coinciding with the vertex of the parabola $C_2$. 

\begin{figure}[t!]
\centerline{\includegraphics[width=0.98\textwidth]{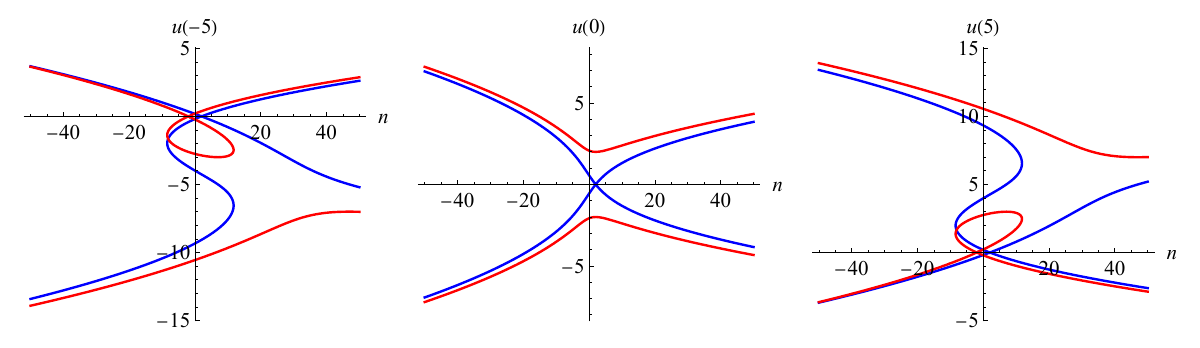}}
\caption{Quasi-stationary curves for dP$_\text{I}$ at $b=0$ and $c=1$, at different values of $t$\\ (blue line: $C^+_2$, red line: $C^-_2$).}
\label{fig:dP1-curve} 
\end{figure}

The derivation of the quasi-stationary curve of the second degree can easily be generalized to the case of the equation with an arbitrary value of the parameter $c$. For this, one has to consider the autonomous maps defined by different formulas at the even and odd points:
\begin{alignat*}{2}
 T^+_n:&\quad u_k(u_{k-1}+u_k+u_{k+1})-2tu_k-\nu+c=0, &&\quad\text{if $k$ is even},\\
 T^-_n:&\quad u_k(u_{k-1}+u_k+u_{k+1})-2tu_k-\nu-c=0, &&\quad\text{if $k$ is odd}.
\end{alignat*}
Finding the fixed points of the composition $T^+_nT^-_n$ reduces to solving the system
\[
 u(2v+u)-2tu-\nu+c=0,\qquad v(2u+v)-2tv-\nu-c=0,
\]
where $u$ and $v$ correspond, respectively, to even and odd values of $k$. The elimination of one of the variables gives an algebraic equation for the other one. As a result, we obtain the equations of two curves, which differ by the sign of $c$:
\begin{equation}\label{curve2pm}
 C^{\pm}_2:~~ 3u^4-8tu^3+2(2t^2+\nu\pm 3c)u^2-(\nu\mp c)^2=0,
\end{equation}
where $C^+_2$ corresponds to even values of $n$ and $C^-_2$ corresponds to odd ones. At $c=0$ these curves merge, the left hand side of (\ref{curve2pm}) factorizes, and we return to the equation (\ref{curve12}). Thus, the introduction of the parameter $c$ leads to a splitting of the union of the parabolas $C_1$ and $C_2$ into a pair of other curves. In Fig.~\ref{fig:dP1-curve} one can see that they exhibit more intriguing rearrangements near the origin, but for large $n$ they asymptotically approach the same parabolas.

Quasi-stationary curves of higher degrees are not defined for dP$_\text{I}$. We restrict ourselves to the consideration of the fixed points of the maps $T^3_n$ and $(T^+_nT^-_n)^2$, but it can be shown that the next powers give nothing new either. At $c=0$, the orbits of $T_n$ of period 3 are described by the system
\[
 u_1(u_3+u_1+u_2)=2tu_1+\nu,\quad
 u_2(u_1+u_2+u_3)=2tu_2+\nu,\quad
 u_3(u_2+u_3+u_1)=2tu_3+\nu,
\]
whence it follows that $u_1=u_2=u_3=\nu/(u_1+u_2+u_3-2t)$. Since all three variables coincide, the solution is exhausted by the fixed points of $T_n$ itself, which we have already taken into account; therefore this solution is discarded. On the contrary, there are too many 4-periodic orbits, so that no definite curve can be assigned to them. In this case we have the system (for an arbitrary $c$)
\begin{alignat*}{2}
 u_1(u_4+u_1+u_2)&=2tu_1+\nu+c,\qquad & u_2(u_1+u_2+u_3)&=2tu_2+\nu-c,\\
 u_3(u_2+u_3+u_4)&=2tu_3+\nu+c,\qquad & u_4(u_3+u_4+u_1)&=2tu_4+\nu-c.
\end{alignat*}
Subtracting the first equation from the third one, and the second one from the fourth one, we obtain
\[
 (u_1-u_3)(u_1+u_2+u_3+u_4-2t)=0,\qquad
 (u_2-u_4)(u_1+u_2+u_3+u_4-2t)=0.
\]
If $u_1+u_2+u_3+u_4\ne 2t$, then $u_1=u_3$, $u_2=u_4$, which gives the 2-periodic solutions already taken into account. Otherwise, one obtains the underdetermined system
\[
 u_1u_3+\nu+c=0,\qquad u_2u_4+\nu-c=0,\qquad u_1+u_2+u_3+u_4=2t.
\]
This agrees with the numerical experiments, which show that, in generic solutions, the variables $u_n$ with the numbers taken modulo 4 lie on separate curves forming characteristic quasiperiodic structures. However, in contrast with the curves $C_1$, $C_2$ and $C^\pm_2$, the shape and position of these curves depend not only on the parameters $b,c,t$, but also on the initial data.

\begin{figure}[t!]
\centerline{\includegraphics[width=0.98\textwidth]{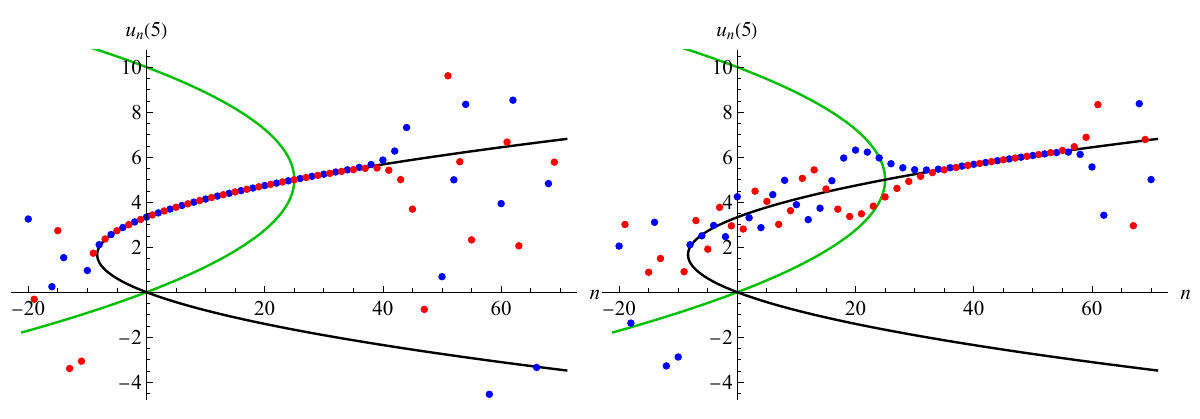}}
\caption{Solutions of dP$_\text{I}(b=c=0,t=5)$ with initial data on the curve $C_1$ at $n_0=20$ (left) and $n_0=45$ (right).}
\label{fig:dP1-plus} 
\end{figure}

\begin{figure}[t!]
\centerline{\includegraphics[width=0.98\textwidth]{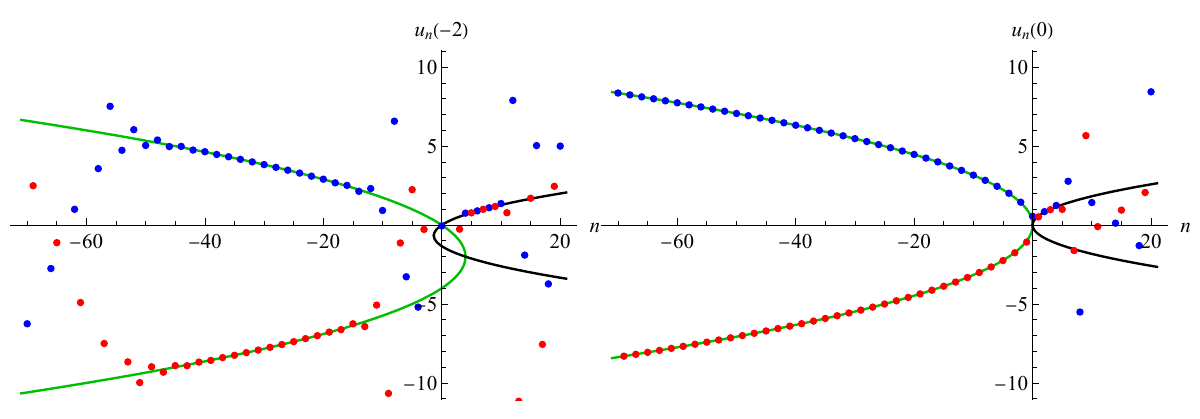}}
\caption{Solutions of dP$_\text{I}(b=c=0)$ with initial data on the curve $C_2$ at $n_0=-30$, for $t=-2$ (left) and $t=0$ (right).}
\label{fig:dP1-minus} 
\end{figure}

Thus, by quasi-stationary solutions of dP$_\text{I}$ we mean solutions asymptotically close, for large $n$, to the curves (\ref{curve2pm}). Although such solutions are special and hidden among the generic ones, they are quite ``friendly'': a solution close to the quasi-stationary curve on a small interval can easily be obtained by solving the Cauchy problem with the initial data at $n_0$, $n_0+1$, taken right on this curve. For the sake of clarity, let us for now restrict ourselves to the case $c=0$. When the initial data are chosen on the upper branch of the parabola $C_1$, the interval of quasi-stationary behavior contains only about 10 points for $t<0$, but it increases for $t>0$ and not too large $n_0$, see the examples in Fig.~\ref{fig:dP1-plus} (in these and subsequent plots, the points with even and odd numbers are shown in blue and red, respectively). Qualitatively, this can be explained by considering the linearized equation (\ref{dP1})
\[
 U_{n-1}+q_nU_n+U_{n+1}=0,\qquad q_n:=1+\nu/u^2_n.
\]
It is stable for $q^2_n<4$, which is equivalent, for $\nu>0$, to the inequality $\nu<u^2_n$. Let the values of $u_n$ correspond to the upper branch of $C_1$; then, for $t<0$, this condition fails at all, while for $t>0$ it selects the part $\nu<t^2$ lying to the left of the intersection point of the parabolas $C_1$ and $C_2$. To the right of this point, the solution gradually loses stability, which is exactly what we observe in the plots presented. Nevertheless, the interval of quasi-stationarity can be increased unboundedly by a more careful choice of the initial data, or by solving a boundary value problem instead of the Cauchy problem.

The examples corresponding to the choice of the initial data on the parabola $C_2$ are shown in Fig.~\ref{fig:dP1-minus}. 
In this case, we observe longer quasi-stationary intervals. At $t=0$, the solution continues along $C_2$ unboundedly far. Besides, one can notice that, when the solution is continued to positive $n$, a small interval of quasi-stationarity on the parabola $C_1$ arises, quite unexpectedly, although this happens only for the integer values of the parameter $b$ (it will be explained later that a switching between the types of asymptotics is possible only if one of parameters $\beta_n$ vanishes, which, at $c=0$, means exactly that $b$ is an integer). This leads to the conjecture that there may exist solutions which are quasi-stationary on the whole axis $n$, that is, asymptotically close to both parabolas, with some transition zone at the origin.

\subsection{Quasi-stationary surfaces}

\begin{figure}[t!]
\centerline{\includegraphics[width=0.98\textwidth]{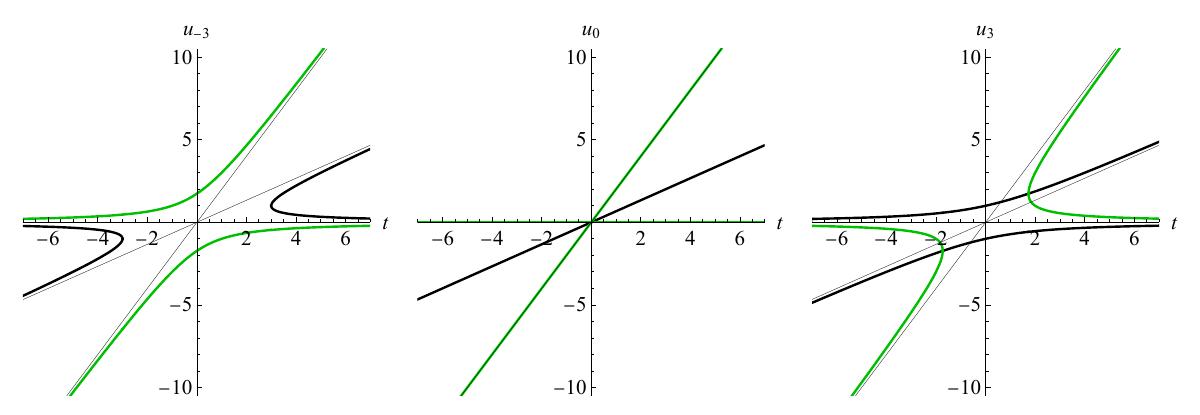}}
\caption{Quasi-stationary curves for P$_\text{IV}$ at $c=0$, for different values of $\nu=n-b$\\
(black line: $C_1$, green line: $C_2$).}
\label{fig:P4-curve-0} 
\end{figure}

\begin{figure}[t!]
\centerline{\includegraphics[width=0.98\textwidth]{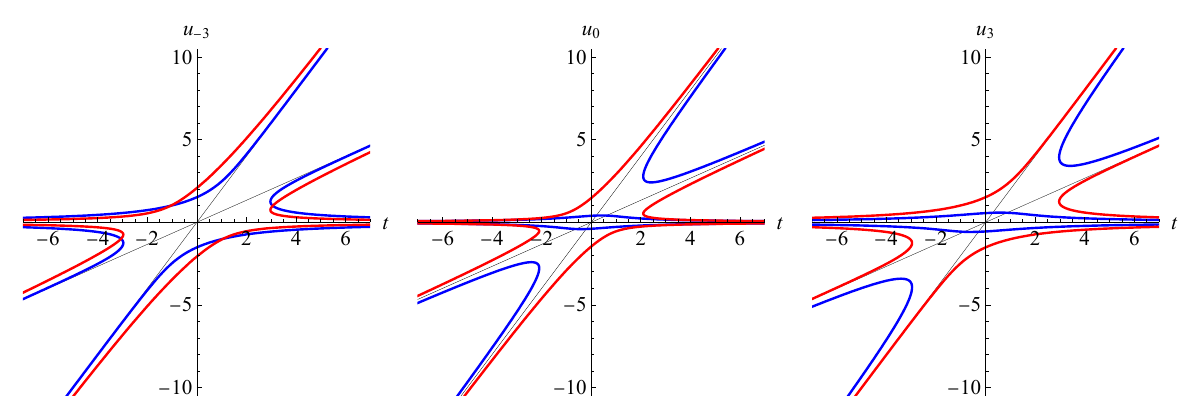}}
\caption{Quasi-stationary curves for P$_\text{IV}$ at $c=1$, for different values of $\nu=n-b$\\ 
(blue line: $C^+_2$, red line: $C^-_2$).}
\label{fig:P4-curve-1} 
\end{figure}

The dependence of quasi-stationary solutions on the parameter $t$ is conveniently studied by using the compatibility of dP$_\text{I}$ with the P$_\text{IV}$ equation (\ref{P4}), which defines the differentiation with respect to $t$. Dropping the terms with derivatives in it, we arrive at the equation 
\begin{equation}\label{P4-curve}
 3u^4-8tu^3+2(2t^2+\alpha)u^2-\beta^2=0,   
\end{equation}
which coincides exactly with the equations of the curves (\ref{curve2pm}), taking into account the formulas (\ref{alphabet}) for the parameters, that is, $\alpha=\nu+3c$, $\beta=\nu-c$ if $n$ is even (the curve $C^+_2$), and $\alpha=\nu-3c$, $\beta=\nu+c$ if $n$ is odd (the curve $C^-_2$). This means that the quasi-stationary curves for P$_\text{IV}$ are given by the same equations as for dP$_\text{I}$, but now we use them to define the dependence on $t$ at a fixed $n$. In the special case $c=0$ (that is, for $\alpha=\beta=\nu$), the P$_\text{IV}$ equation takes the form
\begin{equation}\label{P4-nu}
 2uu''-(u')^2=(3u^2-2tu-\nu)(u^2-2tu+\nu),
\end{equation}
which leads to the equations (\ref{curve1}), (\ref{curve2}), which are now interpreted as the equations of hyperbolas rather than parabolas. The curves for $c=0$ and $c\ne0$ are shown in Fig.~\ref{fig:P4-curve-0} and Fig.~\ref{fig:P4-curve-1}, respectively. In both cases, the same colors are used as in Figs.~\ref{fig:dP1-curve12} and \ref{fig:dP1-curve}.

\begin{figure}[t!]
\centerline{\includegraphics[width=0.75\textwidth]{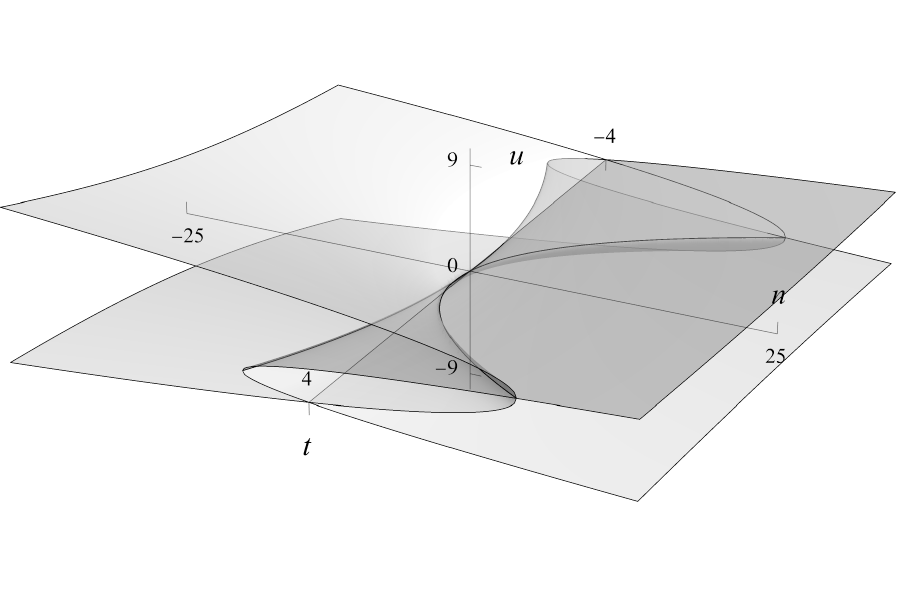}\vspace{-35pt}}
\caption{The common quasi-stationary surface of dP$_\text{I}$ and P$_\text{IV}$ in the case $c=0$.}
\label{fig:dP1-P4_surface} 
\end{figure}

\begin{figure}[t!]
\centerline{\includegraphics[width=0.75\textwidth]{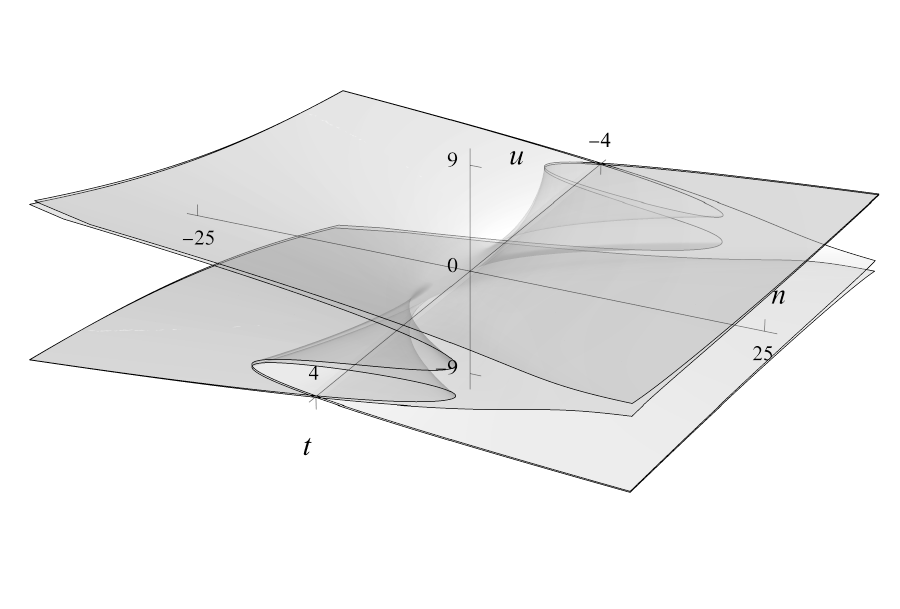}\vspace{-35pt}}
\caption{The common quasi-stationary surface of dP$_\text{I}$ and P$_\text{IV}$ at $c=1/2$.}
\label{fig:dP1-P4_c_surface} 
\end{figure}

Thus, the equations (\ref{curve12}) or (\ref{curve2pm}) can be interpreted not as the equations of curves, but as the equation of the quasi-stationary surface $u=u(n,t)$, common for dP$_\text{I}$ and P$_\text{IV}$, see Figs.~\ref{fig:dP1-P4_surface} and \ref{fig:dP1-P4_c_surface}. Its sections at a fixed $t$ serve as the quasi-stationary curves for dP$_\text{I}$, while the sections at a fixed $n$ serve as the quasi-stationary curves for P$_\text{IV}$. A natural question arises whether there exist common quasi-stationary solutions of dP$_\text{I}$ and P$_\text{IV}$, that is, functions $u_n(t)$ satisfying both equations simultaneously and defining quasi-stationary solutions for each of them. A necessary condition for this is the preservation of the quasi-stationary asymptotics in $n$ under the evolution in $t$, and vice versa. The verification of this property will be performed in Sections \ref{s:t-series} and \ref{s:n-series}.

\begin{remark}
The change $u=U+2t/3$ reduces (\ref{P4-curve}) to the canonical catastrophe equation of the swallowtail type
\[
 U^4+X_1U^2+X_2U+X_3=0
\]
with three control parameters 
\[
 X_1=\frac{2\alpha-4t^2}{3},\quad 
 X_2=\frac{8(3\alpha-2t^2)t}{27},\quad 
 X_3=\frac{8\alpha t^2-9\beta^2}{27}.
\]
\end{remark}

\begin{remark}
In the problem under consideration, the coordinate $n$ is discrete, while $t$ is continuous. This is typical for the standard Painlev\'e equations of the second order, for which the dependence of the parameters on the additional variable is described by discrete B\"acklund transformations (apart from trivial scalings). However, for the higher Painlev\'e equations, the situation when both coordinates are continuous is quite possible. For example, the paper \cite{Suleimanov_1994} studied a fourth order analog of the P$_\text{I}$ equation, depending on a parameter $t$ such that the differentiation with respect to it is governed by the KdV equation. Similar parameters admitting a continuous, rather than discrete, deformation exist in many known Painlev\'e type equations of the fourth order, for example, those from \cite{Cosgrove_2000}. The situation when both coordinates are discrete is also possible.
\end{remark}

\begin{remark}
At $c=0$, the quasi-stationary curves in Figs.~\ref{fig:dP1-curve12} and \ref{fig:P4-curve-0} have, for any $t\ne0$ or $\nu\ne0$, the same shape under an appropriate scaling. Indeed, the passage to the self-similar variables
\begin{equation}\label{uxnu}
 u_n=tu(x),\qquad x=\nu/t^2   
\end{equation}
brings the curve (\ref{curve12}) to a fixed parameter-free form:
\begin{equation}\label{ux}
 (3u^2-2u-x)(u^2-2u+x)=0.
\end{equation}
For $c\ne0$, the scaled curves (\ref{curve2pm}) tend to (\ref{ux}) asymptotically, for large $t$. In Section \ref{s:self}, we use a change of variables close to (\ref{uxnu}) to describe the asymptotics of the quasi-stationary solution in the transition region between the parabolas, which generalizes some results from \cite{Bleher_Its_2003} and \cite{Adler_Suleimanov_2026}.
\end{remark}

\section{Asymptotics in \texorpdfstring{$t$}{t} for a fixed \texorpdfstring{$n$}{n}}\label{s:t-series}

Let $u_n(t)$ be a common solution of dP$_\text{I}$ and P$_\text{IV}$. If it tends to a branch of the quasi-stationary curve, for a fixed $n$, as $t\to+\infty$ or $t\to-\infty$, then it necessarily has a power type asymptotics
\begin{equation}\label{usim}
 u\sim t(p_0+p_1/t+p_2/t^2+\dots)
\end{equation}
with constant coefficients (for generic solutions this is not so, because the coefficients may be oscillating functions, or a solution may have an infinite sequence of poles). The series for $t\to+\infty$ and $t\to-\infty$ may, in general, differ, but at the moment we do not specify which of the limits is considered. A formal solution of the form (\ref{usim}) is constructed in a fairly straightforward way. Substituting it into the P$_\text{IV}$ equation (\ref{P4}) and equating the coefficient of $t^4$ to zero, we obtain
\[
 p^2_0(p_0-2)(3p_0-2)=0,
\] 
which gives three possible values of $p_0$. For all three cases, in the next two orders in $t$ we obtain the equality $p_1=0$ and the expressions for $p_2$; moreover, for $p_0=0$ two values $p_2=\pm\beta/2$ are possible. All subsequent coefficients are determined uniquely, taking into account that, in the case $p_0=\beta=0$, the whole series equals 0, which is easy to prove by keeping track of its leading term upon substitution into the equation. As a result, we arrive at the series
\begin{align}
\label{Aser} 
 A(\alpha,\beta;t)&= 2t -\frac{\alpha}{2t} -\frac{3\alpha^2-\beta^2+4}{16t^3} 
        -\frac{\alpha(9\alpha^2-5\beta^2+44)}{64t^5} +\dots,\\
\label{Bser} 
 B(\alpha,\beta;t)&= \frac{2}{3}t +\frac{\alpha}{2t} +\frac{3\alpha^2-9\beta^2+4}{16t^3} 
        +\frac{3\alpha(3\alpha^2+9\beta^2-28)}{64t^5} +\dots,\\
\label{Cser} 
 C(\alpha,\beta;t)&= \frac{\beta}{2t} -\frac{\beta(\alpha-2\beta)}{8t^3} 
        +\frac{\beta(3\alpha^2-16\alpha\beta+17\beta^2+12)}{64t^5} +\dots,\\
\label{Dser} 
 D(\alpha,\beta;t)&= -\frac{\beta}{2t} +\frac{\beta(\alpha+2\beta)}{8t^3} 
        -\frac{\beta(3\alpha^2+16\alpha\beta+17\beta^2+12)}{64t^5} +\dotso.
\end{align}
Note that these series diverge, except for the cases where they correspond to rational solutions of P$_\text{IV}$, which exist for a discrete set of the parameters $\alpha,\beta$. Indeed, if a series converges to some function of $t$, then $t=\infty$ is a pole or a regular point, and all other singular points lie in a certain bounded region of the complex plane. However, it is known that the general solution of P$_\text{IV}$ is meromorphic, hence the number of such singular points can only be finite, whence it follows that this function is rational. Despite the divergence, the constructed formal series are important as asymptotic expansions for the class of solutions we are interested in. We note the following properties of these series.

\begin{proposition}\label{th:Pid}
All the series found satisfy the identities
\[
  P(\alpha,\beta;-t)=-P(\alpha,\beta;t),\qquad P(-\alpha,-\beta;t)=-iP(\alpha,\beta;it)
\] 
(the first of them means that the series contain only odd powers of $t$). The coefficients of the series $A$ and $B$ contain only even powers of $\beta$, while the series $C$ and $D$ differ by the sign change of $\beta$ and vanish identically for $\beta=0$. 
\end{proposition}
\begin{proof}
The equation (\ref{P4}) is invariant under the changes $u(t)\to-u(-t)$ and $u(t)\to -iu(it)$, $\alpha\to-\alpha$, $\beta\to-\beta$. Applying them to any of the series (\ref{Aser})--(\ref{Dser}), we obtain series of the same type, which are solutions of (\ref{P4}), and the initial coefficients do not change. Since the subsequent coefficients are determined uniquely, the transformed series coincide with the original one. The second statement follows from the invariance of the equation under the change $\beta\to-\beta$.
\end{proof}

\begin{proposition}\label{th:BT} 
Let $A_n$, $B_n$, $C_n$ and $D_n$ be the series (\ref{Aser})--(\ref{Dser}), whose parameters are given according to the formulas (\ref{alphabet}): $\alpha_n=n-b+3(-1)^nc$, $\beta_n=n-b-(-1)^nc$. Then, under the B\"acklund transformations (\ref{BT-gen}), they are transformed as follows:
\begin{alignat*}{2}
 & \text{if~} u_n\sim A_n, &~~& \text{~then~} u_{n\pm1}\sim C_{n\pm1};\\
 & \text{if~} u_n\sim C_n \text{~and~} \beta_n\ne0, &&\text{~then~} u_{n\pm1}\sim A_{n\pm1};\\
 & \text{if~} u_n\sim B_n, && \text{~then~} u_{n\pm1}\sim B_{n\pm1};\\
 & \text{if~} u_n\sim D_n \text{~and~} \beta_n\ne0, &&\text{~then~} u_{n\pm1}\sim D_{n\pm1}.
\end{alignat*}
\end{proposition}
\begin{proof}
The original series for $u_n$ is a formal solution of P$_\text{IV}$ with the parameters $\alpha_n$ and $\beta_n$; here, the condition $\beta_n\ne0$ in the case of the series $C_n$ and $D_n$ means that they do not vanish. It follows that the transformed series corresponding to $u_{n\pm1}$ also satisfies P$_\text{IV}$, with the parameters $\alpha_{n\pm1}$ and $\beta_{n\pm1}$. Since the B\"acklund transformation is given by a rational formula, this series keeps the general structure (\ref{usim}) and hence must belong to one of the four types. To establish the correspondence, it suffices to trace the leading terms of the series.
\end{proof}

The special case $\beta_n=0$ requires a separate consideration, since in this case $C_n=D_n=0$ and the previous proposition is not applicable. If the corresponding solution $u_n(t)$ is indeed identically zero, then the sequence of B\"acklund transformations simply terminates. However, this solution may also be nonzero and merely decay faster than $1/t$, which is not described by the series of the form (\ref{usim}). In this case, the B\"acklund transformation is applicable, and one can trace how the asymptotics of the neighboring solutions $u_{n-1}$ and $u_{n+1}$ are related.

\begin{proposition}\label{th:BT2} 
Let $\beta_n=0$ and the solution $u_n$, which is not identically zero, have the asymptotics $o(t^{-1})$. If, in addition, one of the solutions $u_{n-1}$ or $u_{n+1}$ has the asymptotics of type $A$, then the other one has the asymptotics of type $D$, and vice versa:
\[
 \text{if~~} u_{n-\sigma}\sim A_{n-\sigma}, \text{~then~} u_{n+\sigma}\sim D_{n+\sigma};\qquad
 \text{if~~} u_{n-\sigma}\sim D_{n-\sigma}, \text{~then~} u_{n+\sigma}\sim A_{n+\sigma},
\]
where $\sigma=1$ or $-1$.
\end{proposition}

\begin{proof}
The solutions $u_{n-1},u_n,u_{n+1}$ are related by the dP$_\text{I}$ equation with the parameter $\beta_n=0$ and, by assumption, $u_n\ne0$. Canceling, we obtain the relation
\begin{equation}\label{uuu}
 u_{n-1}+u_n+u_{n+1}=2t,
\end{equation}
where it remains to pass to the series and equate the leading terms, taking into account that the series for $u_n$ is equal to zero. Let us first note that, since $\beta_n=n-b-(-1)^nc=0$, we have
\begin{align*}
 \alpha_{n\pm1}& = n\pm1-b-3(-1)^nc= \pm1-2(-1)^nc,\\ 
 \beta_{n\pm1}& = n\pm1-b+(-1)^nc= \pm1+2(-1)^nc,
\end{align*}
which gives a simple relation between the parameters at the sites $n\pm1$: 
\[
 \alpha_{n\pm1}= -\beta_{n\mp1}.
\]
If $u_{n-1}\sim A_{n-1}= 2t-\alpha_{n-1}/(2t)+\dots$, then from (\ref{uuu}) we have $u_{n+1}\sim\alpha_{n-1}/(2t)=-\beta_{n+1}/(2t)$, and since the series is determined uniquely by its leading coefficient, we conclude that $u_{n+1}\sim D_{n+1}$. The remaining cases are completely analogous.
\end{proof}

Let a solution $u$ have the asymptotics of type $X$ as $t\to-\infty$ and of type $Y$ as $t\to+\infty$, where $X$ and $Y$ are one of the types $\{A,B,C,D\}$ for $\beta\ne0$, or one of the types $\{A,B,0\}$ for $\beta=0$. Then we will say that $u$ has the asymptotics of type $(X,Y)$ and write this as $u\sim(X,Y)$.

Arbitrary combinations of the asymptotics (\ref{Aser})--(\ref{Dser}) are possible, which leads to a large variety of special solutions, see the classification in the paper \cite{Schiff_Twiton_2024}, where the conditions on the parameters under which globally regular solutions with a given asymptotics exist were also studied. However, if we want the common solution of dP$_\text{I}$ and P$_\text{IV}$ to be regular for all $n$, this imposes more rigid restrictions. Indeed, this means that all $u_n$ must have no poles, and no zeros either (except for the case when $u_n=0$ identically), since otherwise $u_{n\pm1}$ will have poles. It follows at once that the asymptotics as $t\to-\infty$ and as $t\to+\infty$ must be of {\em different} types, since otherwise the solution changes sign on the real axis, which is possible only in the presence of a zero or a pole. This also forbids, for example, the combination of the types $A$ and $B$. We will continue the analysis of these restrictions in Sections \ref{s:zero} and \ref{s:regular}.

\section{Boundary value problems}\label{s:bvp}

\subsection{The \texorpdfstring{dP$_\text{I}$}{dP-I} equation}\label{s:dP1-bc}

The numerical method for constructing quasi-stationary solutions of dP$_\text{I}$ consists in solving a boundary value problem on a sufficiently large interval $n_1\le n\le n_2$, with the values at the endpoints taken on the quasi-stationary curve. For positive solutions (for $b+|c|\le n_1$), an explicit iterative scheme is applicable, which consists in that, if $u_n$ is an approximate solution, then the next approximation $\tilde u_n$ is found from the quadratic equation 
\[
 \tilde u_n(u_{n-1}+\tilde u_n+u_{n+1})-2t\tilde u_n-\beta_n=0.
\]
This leads to the map
\begin{equation}\label{dP1-sqrt}
 \tilde u_n= t-\frac{u_{n-1}+u_{n+1}}{2} +\sqrt{\Bigl(t-\frac{u_{n-1}+u_{n+1}}{2}\Bigr)^2+\beta_n},\qquad n_1<n<n_2,
\end{equation}
where $\tilde u_{n_1}=u_{n_1}$, $\tilde u_{n_2}=u_{n_2}$ and the positive value of the root is taken. Its convergence was proved in \cite{Lew_Quarles_1983} (for $t=0$, but the experiments show that it takes place for any $t$). A map formally suitable for constructing solutions taking values of different signs (which is inevitable when considering solutions for $n<0$) can be obtained if, instead of the exact solution of the quadratic equation, one applies one step of the Newton method, which gives, instead of (\ref{dP1-sqrt}), the rational map
\begin{equation}\label{dP1-rat}
 \tilde u_n= \frac{u^2_n+\beta_n}{u_{n-1}+2u_n+u_{n+1}-2t}.
\end{equation}
However, it converges only if the initial approximation is very close to the exact solution. A more robust implicit scheme is obtained if one writes the equation as $E_n=0$, where
\begin{equation}\label{En-gen}
 E_n= u_{n-1}+f_n(u_n)+u_{n+1},
\end{equation}
and determines $\tilde u_n=u_n+\delta_n$ by minimizing $E[u+\delta]$ with respect to the linear approximation. This leads to a linear system for $\delta$ with a tridiagonal matrix:
\begin{equation}\label{delta-gen}
 \delta_{n-1}+f'_n(u_n)\delta_n+\delta_{n+1}=-E_n,\qquad \delta_{n_1}=\delta_{n_2}=0.
\end{equation}
The iterations of the map $u\to u+\delta$ are repeated until the condition $\sum E^2_n<\varepsilon$ holds for a given accuracy $\varepsilon$. For dP$_\text{I}$, the equations take the form
\begin{equation}\label{dP1-delta}
\begin{gathered}
 E_n= u_{n-1}+u_n+u_{n+1}-2t-\frac{\beta_n}{u_n},\qquad \tilde u_n=u_n+\delta_n,\\
 \delta_{n-1}+\left(1+\frac{\beta_n}{u^2_n}\right)\delta_n+\delta_{n+1}=-E_n,\qquad 
 \delta_{n_1}=\delta_{n_2}=0.
\end{gathered}
\end{equation}
As the initial approximation, one can use the sequence 
\[
 u_n = \left\{\begin{aligned} 
   & t+\sigma(-1)^n\sqrt{t^2-n+b} & \text{for}~ n<b+t^2,\\ 
   & \frac{1}{3}(t+\sqrt{t^2+3n-3b}) & \text{for}~ n\ge b+t^2,
 \end{aligned}\right. 
\]
where $\sigma=\pm1$, which corresponds to the points on the parabolas (\ref{curve1}) and (\ref{curve2}). 

This scheme demonstrates excellent convergence for $t\ge0$ and small values of the parameter $c$. The exact restrictions on $c$ are not known; for example, for $b=0.5$ it was found empirically that the values in the interval $-0.246<c<0.5$ are suitable. For $t<0$, the method loses convergence (as well as the map (\ref{dP1-rat})), which is caused by objective reasons, since the solution becomes pole-type, in general.

\begin{figure}[t!]
\centerline{\includegraphics[width=\textwidth]{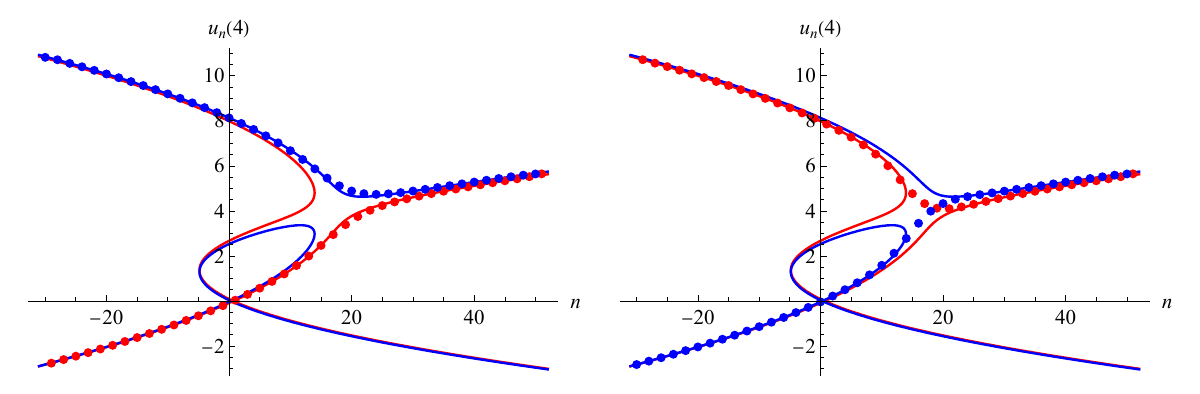}}
\centerline{\includegraphics[width=\textwidth]{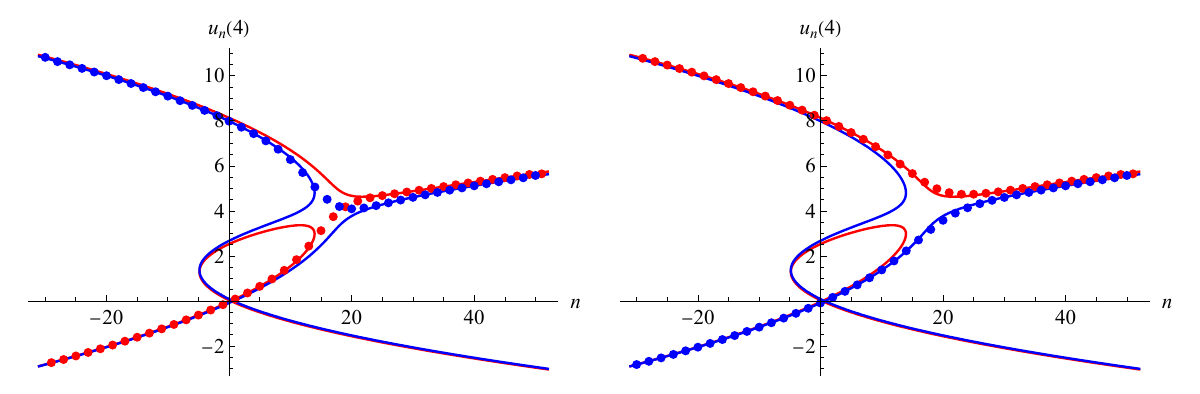}}
\caption{Solutions at $b=1/2$, $c=-1/5$ (top), at $b=1/2$, $c=1/5$ (bottom).}
\label{fig:dP1-bctn} 
\end{figure}

In Fig.~\ref{fig:dP1-bctn}, the plots of the numerical solutions constructed by the scheme (\ref{dP1-delta}) are presented. The top and bottom plots differ in the sign of $c$, which results in the permutation of the quasi-stationary curves $C^+_2$ and $C^-_2$. Since each of these curves consists of two branches, there are also two more variants, which differ in the asymptotics of the points as $n\to-\infty$, depending on their parity. The choice of one or another variant is determined by the sign of $\sigma$ in the initial approximation. In the left plots, the even numbers lie on the upper branch of $C^+_2$, and the odd ones on the lower branch of $C^-_2$; in the right plots, the even ones lie on the lower branch of $C^+_2$, and the odd ones on the upper branch of $C^-_2$. As $n\to+\infty$, both the even and the odd points lie on the upper branches of the corresponding curves.

\begin{remark}
The equations $E_n=0$ define the stationary points of the functional 
\begin{equation}\label{energy}
 \sum_n\Bigl(\frac{1}{2}u^2_n+u_{n-1}u_n-2tu_n-\beta_n\log|u_n|\Bigr),
\end{equation}
and the matrix of the linear system for $\delta_n$ coincides with its Hessian matrix. Under summation over $n>0$ and under the conditions $u_n>0$ and $t\le0$, the a priori estimate $u^2_n<\beta_n$ holds, which makes it possible to prove that the functional is bounded and convex (the Hessian matrix is positive definite). This gives the proof of the existence and uniqueness of the positive solution in the papers \cite{Lew_Quarles_1983, Bonan_Nevai_1984}. However, for $t>0$, as well as for solutions on the whole $n$-axis, this method of proof is not applicable.
\end{remark}

\subsection{The \texorpdfstring{P$_\text{IV}$}{P-IV} equation}\label{s:P4-bc}

For constructing solutions of P$_\text{IV}$, we first make the change $u=\sigma v^2$ to the equation
\begin{equation}\label{P4v}
 v''=\frac{3}{4}v^5-2\sigma tv^3+\left(t^2+\frac{\alpha}{2}\right)v-\frac{\beta^2}{4v^3},
\end{equation}
where the sign $\sigma=\pm1$ is chosen depending on whether we consider a positive or a negative solution $u$. Applying a difference approximation on a uniform grid
\[
 v_k\sim v(t_k),\qquad t_k=t_0+kh,\qquad k=0,\dotsc,K,\qquad h=(t_K-t_0)/K,
\]
we arrive at the equations $E_k=0$, where $k=1,\dotsc,K-1$ and
\begin{equation}\label{P4v-appr}
 E_k= v_{k-1}-2v_k+v_{k+1}+h^2\left(
  -\frac{3}{4}v^5_k+2\sigma t_kv^3_k-\left(t^2_k+\frac{\alpha}{2}\right)v_k
  +\frac{\beta^2}{4v^3_k}\right).
\end{equation}
The solution of this problem is constructed by the general scheme (\ref{En-gen}), (\ref{delta-gen}), according to the formulas
\begin{equation}\label{P4-lin1}
 \tilde v_k=v_k+\delta_k,
\end{equation}
\begin{equation}\label{P4-lin2}
 \delta_{k-1}-\left(2+h^2\left(\frac{15}{4}v^4_k-6\sigma t_kv^2_k+t^2_k+\frac{\alpha}{2}+\frac{3\beta^2}{4v^4_k}\right)\right)\delta_k+\delta_{k+1}=-E_k,\qquad \delta_0=\delta_K=0,
\end{equation}
and the values $v_0$, $v_K$ at the endpoints of the interval and the initial approximation are determined in accordance with the required branch of the quasi-stationary curve (\ref{P4-curve}). 

The described difference scheme works well for the solutions with the asymptotics $u\sim2t$ and $u\sim 0$ (the types $A,C,D$), but it is not suitable for the solutions with the asymptotics $u\sim2t/3$ (type $B$), that is, those approaching the middle branch of the quasi-stationary curve: for them, depending on the choice of the initial approximation, the iterations may converge to an oscillating solution or may not converge at all. However, for the solutions shown in Fig.~\ref{fig:dP1-bctn}, this asymptotics is not needed, since for a fixed $n$ it is intermediate. 

\begin{figure}[t!]
\centerline{\includegraphics[width=\textwidth]{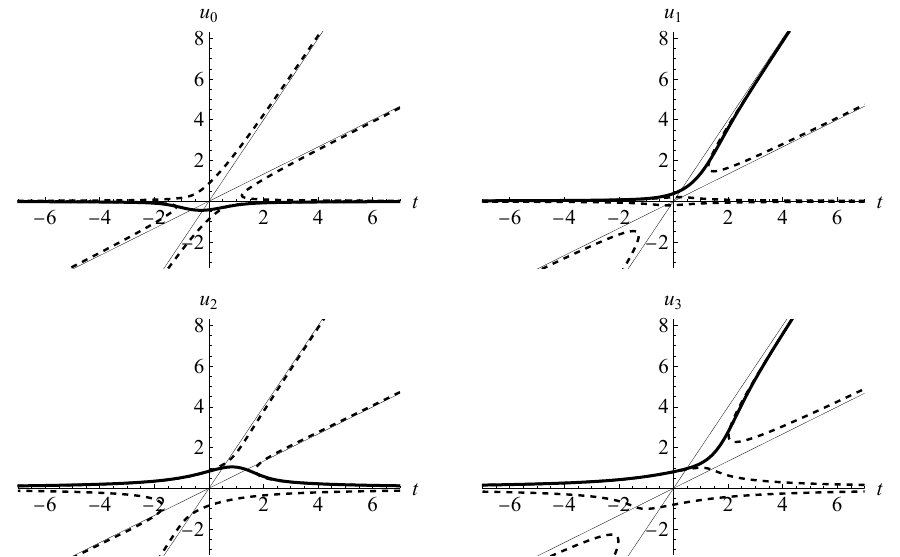}}
\caption{A sequence of quasi-stationary solutions of P$_\text{IV}$ related by the B\"acklund transformation. The dashed lines show the curves (\ref{P4-curve}) for the corresponding values of $\alpha$ and $\beta$.}
\label{fig:P4-DA} 
\end{figure}

Let us illustrate by an example the process of assembling the quasi-stationary surface from solutions of P$_\text{IV}$. Choose some values of the parameters, for example, $b=0.5$ and $c=-0.2$, which corresponds to the top plots in Fig.~\ref{fig:dP1-bctn}. Then, according to the formulas (\ref{alphabet}), the sequence of the parameters in the P$_\text{IV}$ equations is
\[
 \dots~ 
 \alpha_0=-1.1,~\beta_0=-0.3,\quad
 \alpha_1=1.1,~\beta_1=0.3,\quad
 \alpha_2=0.9,~\beta_2=1.7,\quad
 \alpha_3=3.1,~\beta_3=2.3~\dotso.
\]
Assume that $u_1$ tends, as $t\to+\infty$, to the upper branch of the curve $C^-_2$, which corresponds to the top right plot in Fig.~\ref{fig:dP1-bctn}. Then this function has the asymptotics of type $A$ as $t\to+\infty$, and since it must be positive for all $t$ and $\beta_1>0$, the admissible asymptotics as $t\to-\infty$ can only be of type $D$. We denote this as $u_1\sim (D,A)$. Then, according to Proposition \ref{th:BT}, the type of the asymptotics of all $u_n$ is determined by the sequence
\[
 \dots~~ u_0\sim(D,C),\quad u_1\sim (D,A),\quad u_2\sim (D,C),\quad u_3\sim (D,A)~~\dotso,
\]
which gives the required boundary conditions for the difference scheme. Applying it, we obtain the plots in Fig.~\ref{fig:P4-DA}. Note that the function $u_{-1}$ in this example must have a zero, since it has the asymptotics of type $(D,A)$ with the parameter $\beta_{-1}=-1.7<0$, whence it follows that $u_{-1}$ changes sign on the real axis. In this case, the change $u=\sigma v^2$ is not applicable, and the described difference scheme becomes unsuitable. The next function $u_{-2}$ develops a pole. Nevertheless, the solution can be continued to the values $n<0$ by means of the Cauchy problem for dP$_\text{I}$ with the computed $u_0$ and $u_1$.

\section{Simplest exact solutions}\label{s:examples}

It should be emphasized that the preservation of the asymptotics of the form (\ref{Aser})--(\ref{Dser}) under the evolution in $n$ does not yet guarantee that the solution, as a function of $n$, is indeed close to the quasi-stationary curve along its whole length. The point is that the solution may approach the asymptotics in $t$ not uniformly in $n$: this may happen earlier for some $n$ and later for others. This leads to the fact that, at a given value of $t$, the solution may turn out to be close to the quasi-stationary curve only on some interval, outside which the solution may behave like a generic solution with poles. In this section we consider two simple examples related to rational functions and Hermite functions. More solutions of this type can be obtained by applying the full group of B\"acklund transformations (see Remark \ref{rem:BT}). Their study from the point of view of various scenarios of combining the pole and quasi-stationary regimes is of certain interest, despite the fact that all special families of solutions of P$_\text{IV}$ have already been described in great detail in the literature (see, in particular, the papers \cite{Bassom_Clarkson_Hicks_McLeod_1992, Bassom_Clarkson_Hicks_1993, Bassom_Clarkson_Hicks_1995}). 

\begin{figure}[t!]
\centerline{\includegraphics[width=0.5\textwidth]{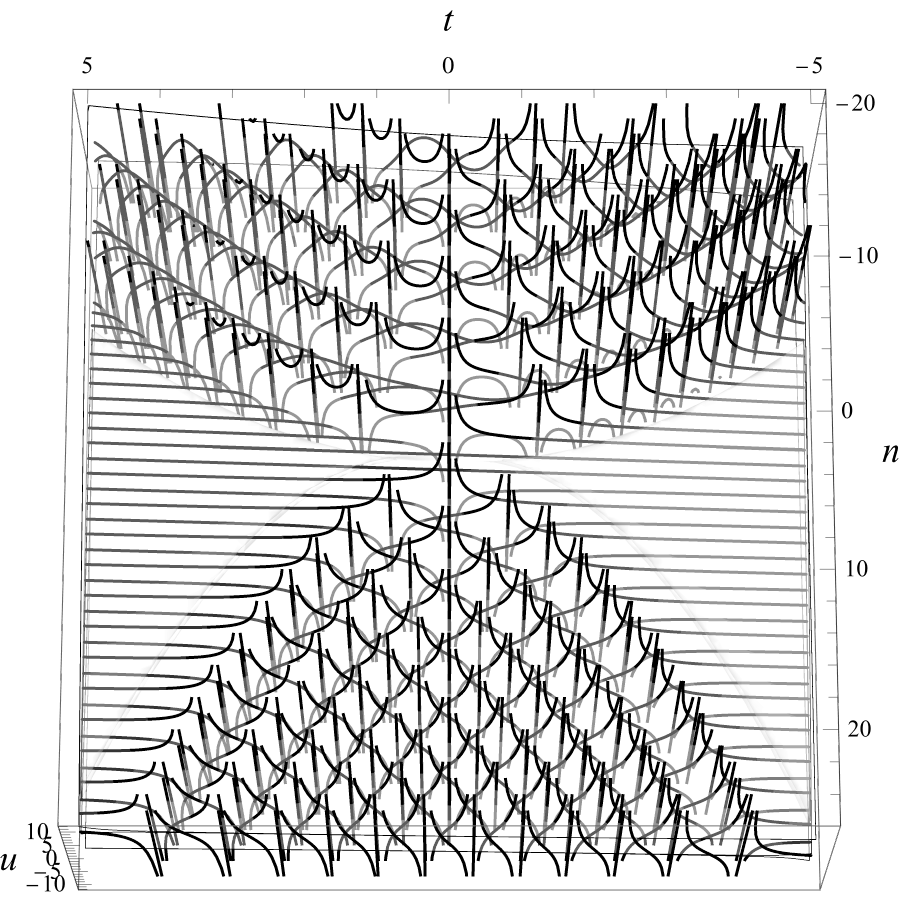}}
\caption{The same solution as a function of $n$ and $t$.}
\label{fig:dP1-23t-3D} 
\end{figure}

\begin{example}\label{ex:23} 
The functions $u_0=u_1=2t/3$ satisfy the system (\ref{P4-sys}) with the parameters $b=1/2$ and $c=1/6$. Using them as the initial data for dP$_\text{I}$, we obtain the sequence
\begin{gather*}
 \dotsc,~
 u_{-3}=\frac{2t(8t^6-60t^4+90t^2-135)}{3(2t^2-3)(4t^4-12t^2-9)},\quad
 u_{-2}=\frac{4t^4-12t^2-9}{3t(2t^2-3)},\quad u_{-1}=\frac{2t^2-3}{3t},\quad u_0=\frac{2t}{3},\\ 
 u_1=\frac{2t}{3},\quad u_2=\frac{2t^2+3}{3t},\quad  u_3=\frac{4t^4+12t^2-9}{3t(2t^2+3)},\quad
 u_4=\frac{2t(8t^6+60t^4+90t^2+135)}{3(2t^2+3)(4t^4+12t^2-9)},~\dots
\end{gather*}
where, according to the formulas (\ref{alphabet}), $u_n$ satisfies P$_\text{IV}$ with the parameters $\alpha_n=n+((-1)^n-1)/2$ and $\beta_n=n-1/2-(-1)^n/6$. 

\begin{figure}[h!]
\centerline{\includegraphics[width=\textwidth]{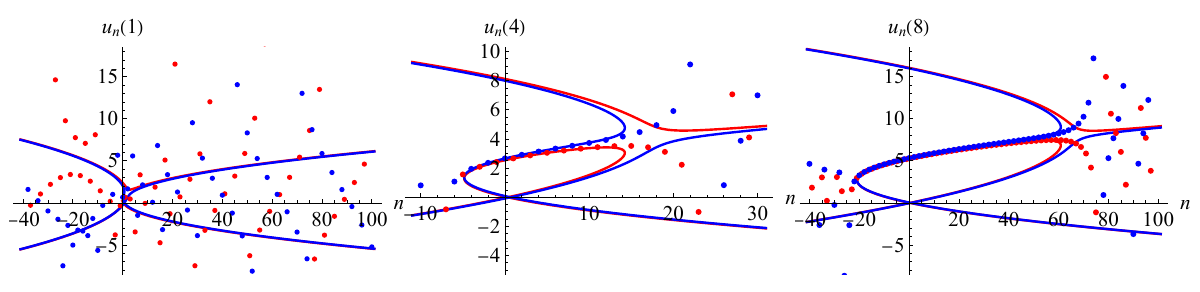}}
\caption{Solution of dP$_\text{I}$ at $b=1/2$, $c=1/6$, $u_0=u_1=2t/3$ at different values of $t$.}
\label{fig:dP1-23t} 
\end{figure}

The plots of this sequence for several fixed values of $t$ are shown in Fig.~\ref{fig:dP1-23t}. In the first plot, at $t=1$, the solution looks like a typical generic solution. Each function $u_n(t)$ (except $u_0$ and $u_1$) has real poles in $t$, but their number is finite, although it grows with $n$. For a fixed $n$, the locations of these poles are confined within a sufficiently large interval in $t$; outside this interval, the solution approaches the asymptotic behavior $2t/3$ which is the same as that of the seed functions $u_0$ and $u_1$. Consequently, as shown in the next two plots, the points $u_n$ settle onto an interval of the quasi-stationary curve growing with $t$, while outside this interval the solution retains its pole-like character. Thus, the solution under consideration is also quasi-stationary in a sense, but not for $n\to\pm\infty$, as we would like, but in the middle. The solution $u_n(t)$ as a function of two variables is shown in Fig.~\ref{fig:dP1-23t-3D}. For this function, the $(n,t)$-plane is divided into two infinite large regions: in one, the function lies close to the quasi-stationary surface, while in the other, it forms a lattice of poles. A conventional boundary between these regions is defined by the projections of the folds of the quasi-stationary surface onto the $(n,t)$-plane (see Fig.~\ref{fig:dP1-P4_c_surface}); these projections are close to the parabolas $3\nu+t^2=0$ and $\nu-t^2=0$.
\end{example}

The following example, as well as the solutions presented in the next section, are related to the Hermite equation
\begin{equation}\label{Hermite}
 y''_\kappa-2ty'_\kappa+2\kappa y_\kappa=0.
\end{equation}
Recall that its solutions satisfy the recurrence relations 
\begin{equation}\label{y-rec}
 y'_\kappa=2\kappa y_{\kappa-1},\qquad y'_{\kappa-1}=2ty_{\kappa-1}-y_\kappa,\qquad
 y_{\kappa+1}-2ty_\kappa+2\kappa y_{\kappa-1}=0,
\end{equation}
which can be considered as linear prototypes for the B\"acklund transformations and the discrete Painlev\'e equations. One of the solutions of the equation (\ref{Hermite}) is the Hermite function defined, for $\operatorname{Re}\kappa<0$, by the formula
\[
 H_\kappa(t)=\frac{1}{\Gamma(-\kappa)}\int^\infty_0e^{-x^2-2tx}x^{-\kappa-1}\,dx.
\]
It also satisfies the recurrence relations (\ref{y-rec}) and is continued to the values $\operatorname{Re}\kappa\ge0$ by means of them. If $\kappa\ne0,1,2,\dotsc$, then the functions $H_\kappa(t)$ and $H_\kappa(-t)$ form a basis of solutions of the equation (\ref{Hermite}). For $\kappa=0,1,2,\dotsc$ one obtains the Hermite polynomials, for which $H_\kappa(t)=(-1)^\kappa H_\kappa(-t)$. In particular, $H_0(t)=1$, and the second linearly independent solution for $\kappa=0$ is given by the imaginary error function $\operatorname{erfi}(t)=-i\operatorname{erf}(it)$. For $\kappa=-1$ we have 
\[
 H_{-1}(t)=\frac{\sqrt{\pi}}{2}e^{t^2}(1-\operatorname{erf}(t)). 
\]
Note also the connection with the parabolic cylinder functions (which occur more often in the reference literature)
\begin{equation}\label{H-D}
 H_\kappa(t)=2^{\kappa/2}e^{t^2/2}D_\kappa(\sqrt{2}t).
\end{equation}

\begin{example}\label{ex:Riccati} 
At $c=1/2$, the dP$_\text{I}$ equation admits the reduction
\begin{equation}\label{d-Riccati}
 u_{2k-1}u_{2k}=2k-2\mu,\qquad u_{2k}+u_{2k+1}=2t,
\end{equation}
where the notation $b+1/2=2\mu$ is introduced. This can be considered as a discrete Riccati equation, since the maps $u_{2k-1}\mapsto u_{2k+1}$ and $u_{2k}\mapsto u_{2k+2}$ are fractional linear. It is checked directly that the reduction (\ref{d-Riccati}) is also compatible with the Volterra lattice, which leads to the continuous Riccati equations defining the reductions of P$_\text{IV}$:
\[
 u'_{2k}=-u^2_{2k}+2tu_{2k}-2k+2\mu,\qquad u'_{2k+1}=u^2_{2k+1}-2tu_{2k+1}+2k+2-2\mu.
\]
Applying the linearizing substitutions, we arrive at the solution
\begin{equation}\label{d-Riccati-sol}
 u_{2k}=\frac{y'_{k-\mu}}{y_{k-\mu}}
  =\frac{2(k-\mu)y_{k-1-\mu}}{y_{k-\mu}},\qquad 
 u_{2k+1}=2t-\frac{y'_{k-\mu}}{y_{k-\mu}}
  =\frac{y_{k+1-\mu}}{y_{k-\mu}},
\end{equation}
where $y_\kappa$ satisfies the Hermite equation (\ref{Hermite}) and the recurrence relations (\ref{y-rec}). Any real solution of the equation (\ref{Hermite}) has zeros for $\kappa>0$ (their number grows linearly with $\kappa$), hence the functions (\ref{d-Riccati-sol}) necessarily have poles for $k>\mu$. However, for $\kappa\le0$ the functions $H_\kappa$ are positive (which follows from the integral representation), and the general solution $y_\kappa$ has at most one zero; moreover, the zeros of all $y_\kappa$ lie in a bounded interval in $t$ (depending on the choice of the solution). It follows that, for $k<\mu$ and $|t|>\operatorname{const}$, the solution (\ref{d-Riccati-sol}) is regular and tends to the quasi-stationary curve, as shown in Fig.~\ref{fig:dP1-Riccati}. In contrast with the previous example, the interval of quasi-stationarity is located on other branches of the curve, corresponding to the asymptotics $u\sim2t$ and $u\sim0$, instead of $u\sim2t/3$. 

\begin{figure}[t!]
\centerline{\includegraphics[width=\textwidth]{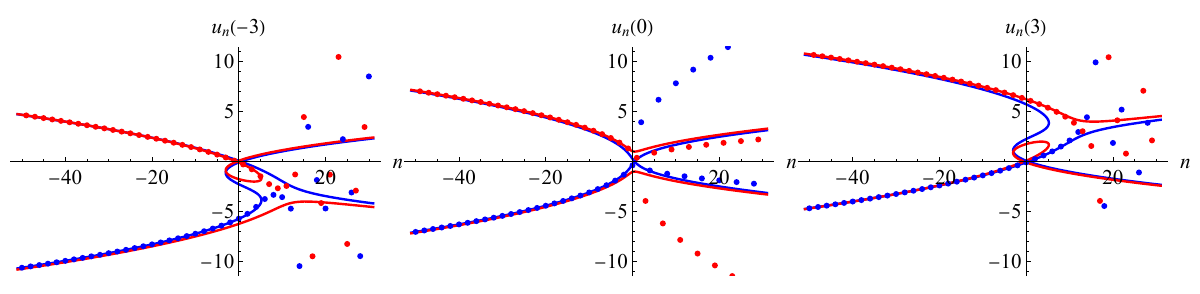}}
\caption{Solution (\ref{d-Riccati-sol}) at $\mu=1/4$ (which corresponds to $b=0$), and
$y_\kappa(t)=H_\kappa(t)$.}
\label{fig:dP1-Riccati} 
\end{figure}

In the solution (\ref{d-Riccati-sol}), the case of integer $\mu$ should be considered separately; it leads to the termination $u_{2\mu}=0$. Up to a shift of the numbering, one can assume that $\mu=0$, $b=-1/2$. Then, restricting to negative numbers, we have $y_k=\gamma_1H_{k}(t)+\gamma_2(-1)^kH_{k}(-t)$,
\[
 u_0=0,\qquad u_{-1}= \frac{2e^{-t^2}}{\sqrt{\pi}(\gamma-\operatorname{erf}(t))},\qquad \gamma=\frac{\gamma_1-\gamma_2}{\gamma_1+\gamma_2}.
\]
If $\gamma_1\gamma_2\ne0$, then some of the $y_k$ have zeros, and the corresponding $u_n$ have poles. For $\gamma_2=0$ we arrive at the solution 
\begin{equation}\label{erf}
 u_{2k}= 2k\frac{H_{k-1}(t)}{H_k(t)},\qquad 
 u_{2k-1}= 2t-\frac{H'_{k-1}(t)}{H_{k-1}(t)} = \frac{H_k(t)}{H_{k-1}(t)},\qquad  
 k=0,-1,-2,\dots    
\end{equation}
(the choice $\gamma_1=0$ gives the same, up to the change $\tilde u_n(t)=-u_n(-t)$). Due to the positivity of $H_k$ with negative indices, the functions $u_n$ are regular for all $t\in\mathbb{R}$ and $n<0$, with $u_{2k}<0$ and $u_{2k-1}>0$. Generalizations of this solution are given in Proposition \ref{th:zero-erf}.
\end{example}

\section{Zero termination}\label{s:zero}

\begin{figure}[h!]
\centerline{\includegraphics[width=0.45\textwidth]{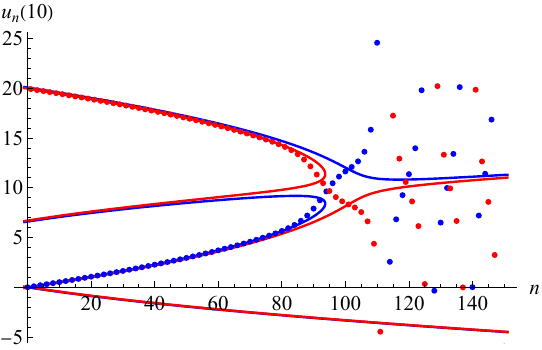}}
\caption{Solution of dP$_\text{I}$ with zero termination. Parameter values: $b=-c=1/3$, $t=10$, the initial value $u_1=Y'/Y$, where $Y=2H_{b-1/2}(t)+H_{b-1/2}(-t)$.}
\label{fig:dP1-zero-generic} 
\end{figure}

Consider solutions with $u_0(t)=0$, which is possible for $c=-b$. In this case, the dP$_\text{I}$ equation and the Volterra lattice split into independent subsystems for positive and negative $n$. The P$_\text{IV}$ equations for the variables $u_{\pm1}$ reduce to the Riccati equations
\begin{equation}\label{Riccati}
 u'_1= -u^2_1 +2tu_1 -2b+1,\qquad u'_{-1}= u^2_{-1} -2tu_{-1} +2b+1,
\end{equation}
which is easily obtained from the system (\ref{P4-sys}) at $n=0$ and $n=-1$. These are the same equations as in Example \ref{ex:Riccati}, but now they hold only for $u_{\pm1}$, not for all $u_n$. Their general solution has the form
\begin{equation}\label{uy}
 u_1= Y'/Y,\qquad u_{-1}=2t-Z'/Z,
\end{equation}
where $Y$ and $Z$ are arbitrary solutions of the Hermite equation (\ref{Hermite}) with the parameter $\kappa=b-1/2$: 
\begin{equation}\label{y-eq}
 y''-2ty'+(2b-1)y=0.
\end{equation}
The remaining functions $u_n$ are expressed through $Y$ or $Z$ by means of the Hankel determinants, whose order grows as $|n|/2$, that is, in this case the answer becomes much more complicated compared with the solution (\ref{d-Riccati-sol}). However, we will not need these formulas. From the known asymptotic behavior for the Hermite functions, or directly from the Riccati equations (\ref{Riccati}), it follows that the functions $u_{\pm1}$ tend to $0$ or $2t$ as $t\to\pm\infty$ (depending on the choice of the solution) and have the asymptotics of type $A$, $C$ or $D$ from Section \ref{s:t-series}. Then, according to Proposition \ref{th:BT}, the functions $u_n$ constructed from them by means of the dP$_\text{I}$ equation also have an analogous asymptotics. Consequently, for large $t$, the solution, viewed as a function of $n$, condenses onto the quasi-stationary curve, but, in general, not uniformly in $n$. Thus, in spite of the complication of the formulas, the solutions of this type behave exactly in the same way as the solutions of Example \ref{ex:Riccati}, the only difference being that now they are restricted to a half-line. A typical example for $n>0$ is shown in Fig.~\ref{fig:dP1-zero-generic}. A more interesting question is how to single out, from the family under consideration, the solutions which are regular in $t$. Below we consider positive and negative $n$ separately.

\subsection{Solutions for \texorpdfstring{$n>0$}{n>0}}\label{s:zero+}

The following proposition was proved in the papers \cite{Lew_Quarles_1983, Bonan_Nevai_1984} for $t\le0$ and \cite{Clarkson_Jordaan_Kelil_2015, Clarkson_Jordaan_2018} for all $t$ (note that these papers use the parabolic cylinder functions, instead of the Hermite functions, related by the formula (\ref{H-D})). 

\begin{proposition}\label{th:zero+}
For $c=-b$, $u_0=0$ and an arbitrary $t\in\mathbb{R}$, the solution of dP$_\text{I}$, positive for all $n>0$, exists if and only if $b<1/2$. This solution is unique and corresponds to the initial condition
\begin{equation}\label{u1(t)}
 u_1(t)= -\frac{H'_{b-\frac{1}{2}}(-t)}{H_{b-\frac{1}{2}}(-t)}
  = (1-2b)\frac{H_{b-\frac{3}{2}}(-t)}{H_{b-\frac{1}{2}}(-t)}.
\end{equation}
In particular, at $t=0$
\begin{equation}\label{u1(0)}
 u_1(0)=2\frac{\Gamma\left(\frac{3}{4}-\frac{b}{2}\right)}
 {\Gamma\left(\frac{1}{4}-\frac{b}{2}\right)}.
\end{equation}
\end{proposition} 

Note that the formulation of this statement does not, a priori, assume either that a positive solution to the dP$_\text{I}$ is compatible with P$_\text{IV}$, or that $u_1$ must satisfy the Riccati equation (\ref{Riccati}); on the contrary, it is proven that these properties hold automatically. For $t\le0$, the existence and uniqueness of the solution can be proven by using the boundedness and convexity of the energy type functional (\ref{energy}). For $t>0$, the papers \cite{Clarkson_Jordaan_Kelil_2015, Clarkson_Jordaan_2018} use more specialized arguments based on Favard's theorem on the relation between the weight for orthogonal polynomials and the coefficients of the three-term recurrence relation. This makes it possible to prove that the evolution of the positive solution of dP$_\text{I}$ governed by the Volterra lattice is well defined and preserves positivity, which reduces the uniqueness question to the case $t=0$.  Some generalizations of this result and a detailed review of the literature on this subject can be found in \cite{Alsulami_Nevai_Szabados_Assche_2015, Clarkson_Loureiro_Assche_2016}. In particular, the paper \cite{Clarkson_Loureiro_Assche_2016} provides a formula defining the unique positive solution of the so-called alternative dP$_{\text{I}}$ equation (also with zero termination condition). It is analogous to the formula (\ref{u1(t)}), with the Hermite function replaced by the Airy function, which defines special solutions of the P$_\text{II}$ equation instead of P$_\text{IV}$. Similar results for one more version of dP$_{\text{I}}$, related to P$_\text{V}$, are obtained in \cite{Clarkson_Dzhamay_Hone_Mitchell_2025}, see also \cite{Aptekarev_Novokshenov_2024}.

The quasi-stationarity of the positive solution at $t=0$ follows from the results of the paper \cite{Lew_Quarles_1983}, where it was shown that $u_n\sim\sqrt{n/3}$ as $n\to\infty$, and several terms of the asymptotic series in inverse powers of $\sqrt{n}$ were derived. In Section \ref{s:n-series} we construct an analogous expansion for an arbitrary $t$, with the leading term of the asymptotics $\frac{1}{3}\sqrt{3n-3b+t^2}$, and show that it keeps its form under the dynamics governed by the Volterra lattice. Hence follows the quasi-stationarity of the solution as $n\to\infty$ at any value of $t$. At the same time, for each fixed $n$, the behavior of $u_n(t)$ is determined by the series of type $A$ (\ref{Aser}) or $C$ (\ref{Cser}) as $t\to+\infty$, or of type $D$ (\ref{Dser}) as $t\to-\infty$, which corresponds to the transition of the solution to one of the branches of the quasi-stationary curve.

\begin{figure}[t!]
\centerline{\includegraphics[width=\textwidth]{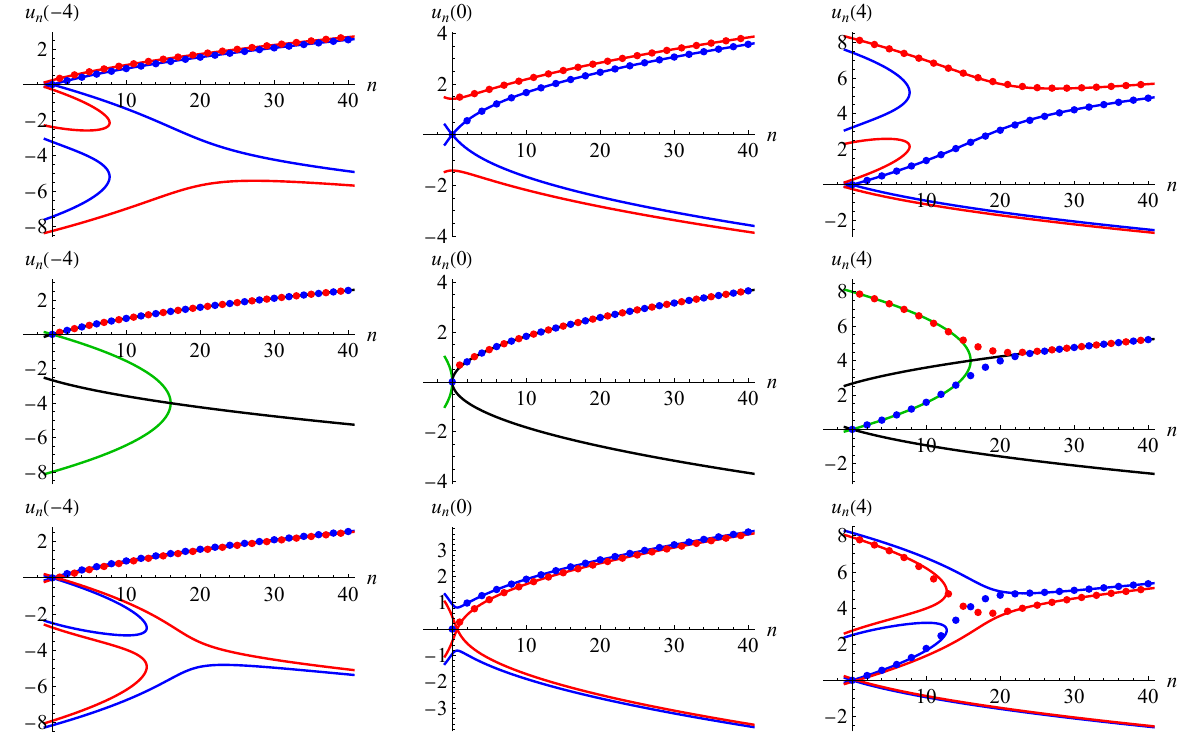}}
\caption{Solution of dP$_\text{I}$ with zero termination and the initial data $u_1(t)$ given by the formula (\ref{u1(t)}). Top row: $b=-1$, middle row: $b=0$, bottom row: $b=1/3$.}
\label{fig:dP1-zero-plus} 
\end{figure}

The profiles of the solution for several values of $t$ and $b$ are shown in Fig.~\ref{fig:dP1-zero-plus}. The middle row corresponds to the value $b=-c=0$, for which the quasi-stationary curve is the union of two parabolas. When $b$ passes through 0, the curves $C^+_2$ (blue line) and $C^-_2$ (red line) swap places. Since the variables $u_n$ with even numbers tend to $C^+_2$, and those with odd numbers tend to $C^-_2$, this leads, for $0<b<1/2$, to the formation of a twist in the last of the plots. 

\begin{remark}\label{rem:precision}
The construction of quasi-stationary solutions by solving the Cauchy problem requires extremely high accuracy in specifying the initial data. Rounding errors lead to the fact that, instead of a quasi-stationary solution, we obtain a pole-type one, for $n$ sufficiently large. For example, in order to construct the famous solution of the equation with $b=t=0$ (the original Shohat's case \cite{Shohat_1939}) up to the point $n=40$, as in the central plot of Fig.~\ref{fig:dP1-zero-plus}, one has to take no less than 25 decimal digits of the number   
\[
 u_1(0)=\frac{H_{-\frac{3}{2}}(0)}{H_{-\frac{1}{2}}(0)}
  =2\frac{\Gamma\left(\frac{3}{4}\right)}{\Gamma\left(\frac{1}{4}\right)}
  \approx 0.67597824006728472899544768467080574828728345491541\dots
\]
Here 50 digits are written out, which is sufficient to continue the solution approximately to $n=80$, and so on, in the same proportion, that is, 500 digits are required for $n=800$. Approximately the same accuracy is required for the general formula (\ref{u1(t)}) as well. In fact, this is not a problem for modern symbolic computation systems, in which both the algorithms for computing the values of special functions and the algorithms for arbitrary precision arithmetic are implemented (we used {\sl Mathematica} \cite{Mathematica_2021}). Of course, theoretically, this is not very satisfactory. As the authors of the article \cite{Clarkson_Loureiro_Assche_2016} rightly point out: ``it is not wise to compute this positive solution starting from the initial values''. Instead of the Cauchy problem, one should solve a boundary value problem, applying, for example, the iterative scheme (\ref{dP1-sqrt}). On the other hand, the Cauchy problem is a more flexible tool, which makes it possible to study also the pole-type solutions such as those shown in Figs.~\ref{fig:dP1-23t} and \ref{fig:dP1-zero-generic}; boundary value problems are not suitable for them.
\end{remark}

\subsection{Solutions for \texorpdfstring{$n<0$}{n<0}}\label{s:zero-}

For $n<0$, instead of the positivity property of the solution one should consider the alternation of signs, but this property is no longer the determining one. For example, at $b=c=t=0$ and $u_0=0$, {\em any} solution of dP$_\text{I}$ for $n<0$ is sign-alternating, whereas for $n>0$ the positive solution is unique. Singling out the sign-alternating solutions for arbitrary parameters is a very difficult problem. We will radically simplify it by requiring, in addition, that the evolution in $t$ governed by P$_\text{IV}$ preserve the signs of $u_n(t)$ for all $t$, which is equivalent to the regularity of the solution for all $t$. This is a narrower setting than in Proposition \ref{th:zero+}, where the compatibility of the positive solution with P$_\text{IV}$ is proved. All regular solutions are described in the following proposition. This family corresponds to a discrete set of parameters and, at $m=0$, coincides with the solution (\ref{erf}). 

\begin{proposition}\label{th:zero-erf}
For $c=-b$ and $u_0=0$, a common solution of dP$_\text{I}$ and P$_\text{IV}$, regular for $t\in\mathbb{R}$ for all $n<0$, exists only for negative half-integer $b$:
\begin{equation}\label{bm}
 b= m-1/2,\qquad m=0,-1,-2,\dotso.
\end{equation}
Up to the change $\tilde u_n(t)=-u_n(-t)$, this solution is unique, is determined by the initial value
\begin{equation}\label{u-1m}
 u_{-1}(t)= \frac{H_m(t)}{H_{m-1}(t)} 
\end{equation}
and satisfies the inequalities $u_{2k}<0$ and $u_{2k+1}>0$.
\end{proposition}

\begin{proof}
For $c=-b$, the parameters $\beta_n$ have the form
\[
 \beta_{2k+1}=2k+1-2b,\qquad \beta_{2k}=2k.
\]
If the condition (\ref{bm}) does not hold, then $\beta_n\ne0$ for all $n<0$. We show that regular solutions are impossible in this case.

The common solution of dP$_\text{I}$ and P$_\text{IV}$ is constructed from the function $u_{-1}$ satisfying the second of the equations (\ref{Riccati}). Its asymptotics can have the form $u_{-1}\sim 2t$ (type $A$ (\ref{Aser})) or $u_{-1}\sim(1+2b)/(2t)=-\beta_{-1}/(2t)$ (type $D$ (\ref{Dser})). If the asymptotics as $t\to-\infty$ and as $t\to+\infty$ are of the same type, then $u_{-1}$ changes sign on the real axis. This means that either $u_{-1}$ vanishes at some $t$, and then $u_{-2}$ has a pole at this point, or $u_{-1}$ itself has a pole. Thus, for the regularity of the solution it is necessary that the asymptotics be different. Taking into account the change $\tilde u_n(t)=-u_n(-t)$, it suffices to consider the combination of the types $D$ as $t\to-\infty$ and $A$ as $t\to+\infty$. Then, according to Proposition \ref{th:BT} under the condition $\beta_n\ne0$, the asymptotics of all functions $u_n$ have the following form:
\begin{equation}\label{DADC}
\begin{aligned}
 (D,A):\qquad & 
 u_{2k+1}\sim -\frac{\beta_{2k+1}}{2t} \text{~~as~~} t\to-\infty, &\qquad & 
 u_{2k+1}\sim 2t \text{~~as~~} t\to+\infty,\\
 (D,C):\qquad &
 u_{2k}\sim -\frac{\beta_{2k}}{2t} \text{~~as~~} t\to-\infty, &\qquad & 
 u_{2k}\sim \frac{\beta_{2k}}{2t} \text{~~as~~} t\to+\infty,
\end{aligned}
\end{equation}
where $k=-1,-2,\dotso$. But, for $k\ll0$, the coefficient $\beta_{2k+1}=2k+1-2b$ is negative, whence it follows that $u_{2k+1}$ changes sign, which means the appearance of a pole either in this function itself or in $u_{2k}$. 

In what follows we assume that (\ref{bm}) holds; then $\beta_{2m-1}=0$. Let us trace the asymptotics of $u_n$ in this case. For $n>2m-1$, it is still determined by the formulas (\ref{DADC}). The function $u_{2m-1}$ has the asymptotics of type $(0,A)$, hence it is positive as $t\to+\infty$ and does not vanish identically. Then, according to Proposition \ref{th:BT2}, when passing through this number, the asymptotics as $t\to-\infty$ switches from type $D$ to type $A$. As a result, for $n<2m-1$, the functions $u_n$ have the asymptotics of type $(A,C)$ if $n$ is even, and of type $(C,A)$ if $n$ is odd. It follows that $u_n$ has the same sign as $t\to\pm\infty$ for any $n<0$, except, possibly, $n=2m-1$, and the previous argument forbidding regular solutions loses its force. 

The general solution of the Riccati equation for $u_{-1}$ has the form 
\[
 u_{-1}=\frac{\gamma_1H_m(t)-\gamma_2H_m(-t)}{\gamma_1H_{m-1}(t)+\gamma_2H_{m-1}(-t)}.
\]
The functions $H_\kappa(t)$ with $\kappa<0$ are positive, grow as $t\to-\infty$ and tend to 0 as $t\to+\infty$. It follows that if the constants $\gamma_1$ and $\gamma_2$ have different signs, then the denominator of the fraction vanishes and $u_{-1}$ has a pole. If $\gamma_1$ and $\gamma_2$ have the same sign and $m\ne0$, then the numerator vanishes; then $u_{-2}$ has a pole. The case $m=0$ was considered in Example \ref{ex:Riccati}. Consequently, for the regularity of the solutions it is necessary that one of the constants vanish, which leads to the function (\ref{u-1m}), up to the indicated change. 

Since the function (\ref{u-1m}) has the asymptotics of type $(D,A)$, it follows from the above analysis that $u_{2k}<0$ as $t\to\pm\infty$ and $u_{2k+1}>0$ as $t\to\pm\infty$, except for the function $u_{2m-1}$, about which it is only known that it is positive as $t\to+\infty$. To complete the proof, it remains to show that all $u_n$ are regular and have no zeros. For $u_{-1}$ this is true. Assume that this is true for some $u_{n+1}$, with $n<-1$, and prove the statement for $u_n$. For this purpose we use the system (\ref{P4-sys}):
\[
 u'_n= u_n(2u_{n+1}+u_n)-2tu_n-\beta_n,\qquad
 u'_{n+1}= -u_{n+1}(u_{n+1}+2u_n)+2tu_{n+1}+\beta_{n+1}.
\]
The function $u_n$ is expressed through $u_{n+1}$ from the second equation by division; since $u_{n+1}$ has no zeros, $u_n$ is regular. From the first equation we obtain that, at the zeros of $u_n$, its derivative equals $-\beta_n$. If $\beta_n\ne0$, then it follows that $u_n$ has at most one simple zero; but, as we have seen, $u_n$ has the same sign as $t\to\pm\infty$, which means that there are no zeros. If $\beta_n=0$ (at $n=2m-1$) and $u_n$ has a zero, then $u_n=0$ identically, which contradicts its positive asymptotics as $t\to+\infty$. Thus, in this case there are no zeros either. By induction, we conclude that all $u_n$ are regular and have no zeros.
\end{proof}

\begin{figure}[t!]
\centerline{\includegraphics[width=\textwidth]{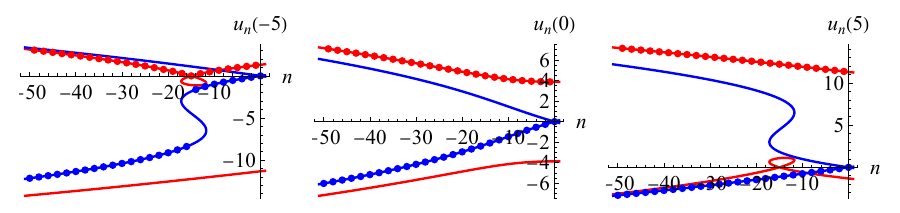}}
\caption{Solution of dP$_\text{I}$ for $n<0$, with the initial data $u_0=0$, $u_{-1}=H_{-7}(t)/H_{-8}(t)$.}
\label{fig:dP1-zero-minus-7} 
\end{figure}

An example of the solution with the initial data of the form (\ref{u-1m}), for $m=-7$, is illustrated in Fig.~\ref{fig:dP1-zero-minus-7}. For $t<0$, the switching of the asymptotics from type $D$ to the alternation of the types $A$ and $C$ is clearly seen when passing through the point $n=2m-1$.

\begin{figure}[t!]
\centerline{\includegraphics[width=0.8\textwidth]{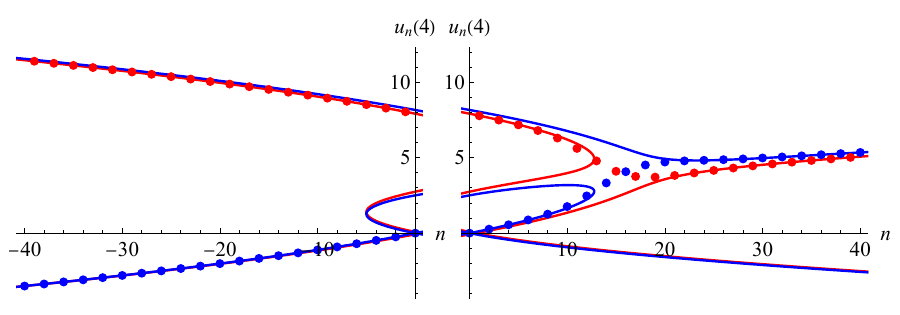}}
\caption{Plots of the solutions with the initial data (\ref{u1(t)}) and (\ref{u-1b}), at $b=1/3$.}
\label{fig:dP1-zero-pm} 
\end{figure}

As is clear from the above proof, the restriction (\ref{bm}) is related to the fact that, if $b$ is not half-integer, then poles appear in the solution. However, it turns out that they are all located on the half-line $t<0$, that is, the solution with the initial value
\begin{equation}\label{u-1b}
 u_{-1}(t)= \frac{H_{b+1/2}(t)}{H_{b-1/2}(t)}, 
\end{equation}
for an arbitrary value $b<1/2$, preserves regularity at least for $t\ge0$. The proof of this fact is rather complicated and goes beyond the scope of our paper. It can be obtained on the basis of the theory of orthogonal polynomials and amounts to proving that, at $t=0$, the inequalities $u_{2k}(0)<0$ and $u_{2k+1}(0)>0$ hold, whence, by arguments analogous to the proof of Proposition \ref{th:zero-erf}, it is deduced that the functions keep the same sign for $t>0$ as well.

The solutions with the initial data (\ref{u1(t)}) and (\ref{u-1b}) are complementary to each other, in the sense that the points with positive and negative numbers lie, for $t>0$, on the continuations of the same branches of the quasi-stationary curve, see Fig.~\ref{fig:dP1-zero-pm}. In particular, note that the formula (\ref{u-1b}) can be written in the equivalent form as
\[
 u_{-1}(t)= 2t+(1-2b)\frac{H_{b-3/2}(t)}{H_{b-1/2}(t)}, 
\]
and the comparison with (\ref{u1(0)}) shows that $u_{-1}(0)=u_1(0)=2\Gamma\left(\frac{3}{4}-\frac{b}{2}\right)/\Gamma\left(\frac{1}{4}-\frac{b}{2}\right)$.

\section{Globally regular solutions}\label{s:regular}

Consider the question of common quasi-stationary solutions of dP$_\text{I}$--P$_\text{IV}$, regular in $t$ for all $n$, assuming that none of the functions $u_n$ vanishes identically. In such a solution, all $u_n$ must have neither zeros nor poles. We will call such solutions globally regular. Generalizing the argument from the proof of Proposition \ref{th:zero-erf}, we arrive at the following necessary condition.

\begin{proposition}
A globally regular quasi-stationary solution without zero termination is possible only if, for some number $n=k$, the following conditions hold: $\beta_{k-1}<0$, $\beta_k=0$ and $\beta_{k+2m+1}\ne0$ for all $m$. Up to a shift of the numbering sending $k$ to 0, this is equivalent to $-c=b>-1/2$ and $b$ not being a half-integer. 
\end{proposition}
\begin{proof}
Assume that all $\beta_n\ne0$. Each of the functions $u_n$ has the asymptotics of one of the types (\ref{Aser})--(\ref{Dser}), which determines its signs as $t\to-\infty$ and $t\to+\infty$. The sign preservation condition allows only the following pairs of types of asymptotics, up to a permutation (which is immaterial due to the change $\tilde u_n(t)=-u_n(-t)$): 
\[
 \begin{array}{llll}
 (D,A), &(D,B), &(D,C) &\text{~for~~}\beta_n>0,\\ 
 (C,A), &(C,B), &(D,C) &\text{~for~~}\beta_n<0.
\end{array}
\]
Proposition \ref{th:BT} makes it possible to trace the passage to the neighboring functions $u_{n\pm1}$. The pair $(C,B)$ goes into the forbidden pair $(A,B)$, hence it should also be discarded. The pair $(D,B)$ goes into the pair $(D,B)$. Then all $u_n$ must have the asymptotics of the same type, but, since $\beta_n<0$ for $n\ll 0$, for such $n$ the pair $(D,B)$ becomes forbidden. The pair $(D,A)$ goes into $(D,C)$, $(D,C)$ goes into $(D,A)$, and $(C,A)$ goes into $(A,C)$. Thus, in the sequence of pairs there is always a subsequence of pairs $(D,A)$ or $(C,A)$. The first becomes forbidden for $n\ll0$, and the second for $n\gg0$. Thus, starting from an arbitrary pair of asymptotics, we arrive, for sufficiently large negative or positive $n$, at a pair in which the sign preservation is violated.    

Thus, one must have $\beta_k=0$ for some $k$. Since it is assumed that $u_k\ne0$, the dP$_\text{I}$ equation is solved at this point as
\[
 u_{k-1}+u_k+u_{k+1}=2t.
\]
Let us trace the asymptotics of the functions involved here. Up to the change $\tilde u_n(t)=-u_n(-t)$, one can assume that $u_k$ has the zero asymptotics as $t\to-\infty$; then, as $t\to+\infty$, its asymptotics can be of type $A$ or $B$. According to Propositions \ref{th:BT} and \ref{th:BT2}, the following cases are possible:
\[
 \begin{array}{ccc}
  u_{k-1} & u_k & u_{k+1} \\
 \hline    
 (A,C) & (0,A) & (D,C) \\
 (D,C) & (0,A) & (A,C) \\
 (A,B) & (0,B) & (D,B) \\
 (D,B) & (0,B) & (A,B) \\
 \end{array}
\]
The last two cases are discarded, since they involve the forbidden pair $(A,B)$. Thus, the asymptotics of type $B$ cannot appear at all in a globally regular solution. 

In the second case, if $\beta_{k-1}=0$, then $u_{k-1}\sim(0,0)$, which means that $u_{k-1}=0$ identically, contrary to the assumption. If $\beta_{k-1}\ne0$, then the continuation of the solution to $u_{k-2}$ gives $u_{k-2}\sim(D,A)$, which is forbidden, since $\beta_{k-2}=-2<0$. 

Only the first case remains. Consider the continuation of the solution to the numbers $n<k$. If also $\beta_{k-1}=0$, then $\beta_{k-2}=\beta_{k-3}=-2$, and
\[
 u_{k-3}\sim(A,D),\quad u_{k-2}\sim(C,D),\quad u_{k-1}\sim(A,0),\quad u_k\sim(0,A),
\]  
with the forbidden pair $(A,D)$. If $\beta_{k-1}\ne0$, then $u_{k-1}\sim(A,C)$, which leads to the restriction $\beta_{k-1}<0$. Let it hold; then $\beta_n<0$ for all $n<k$ and the sequence arises
\[
 \dots~~ u_{k-4}\sim(C,A),\quad u_{k-3}\sim(A,C),\quad u_{k-2}\sim(C,A),\quad u_{k-1}\sim(A,C), 
\]
in which all pairs are allowed. The continuation of the solution to the numbers $n>k$ gives, in the case when all $\beta_n\ne0$, the sequence
\[
 u_{k+1}\sim(D,C),\quad u_{k+2}\sim(D,A),\quad u_{k+3}\sim(D,C),\quad u_{k+4}\sim(D,A),~~\dotsc,
\]
in which all pairs are allowed, since $\beta_{k+2m}=2m>0$. If, however, one of the parameters $\beta_{k+2m+1}$ vanishes, then the corresponding function $u_{k+2m+1}$ has the asymptotics of type $(0,0)$ and vanishes identically, which corresponds to the sequence with zero termination, contrary to the assumption. It is not difficult to see that this is exactly the same case as considered in Proposition \ref{th:zero-erf}, up to a shift of the numbering which places the zero termination at the point $n=k+2m+1$ into the point $n=0$.

The shift of the numbering allows one to set $k=0$ without loss of generality; then the restrictions take the form $\beta_{-1}=-1-b+c<0$, $\beta_0=-b-c=0$, which gives the indicated inequalities.
\end{proof}

\begin{figure}[t!]
\centerline{\includegraphics[width=\textwidth]{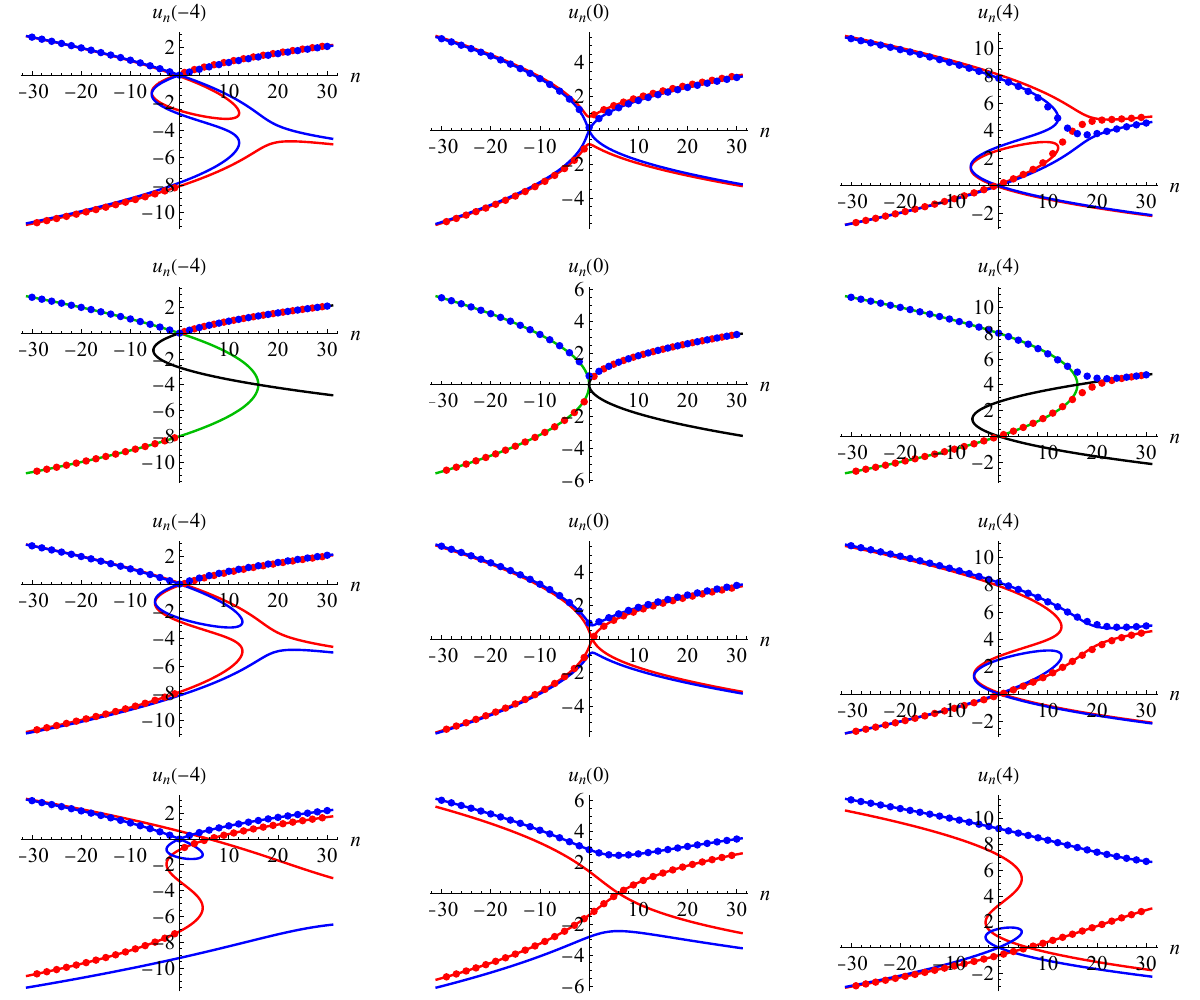}}
\caption{Globally regular solutions of dP$_\text{I}$--P$_\text{IV}$ at the parameter values\\ 
$-c=b=-1/3$, $0$, $1/3$, $3$ (top to bottom).}
\label{fig:dP1-P4-global} 
\end{figure}

\begin{figure}[t!]
\vspace{-25pt}\centerline{\includegraphics[width=0.7\textwidth]{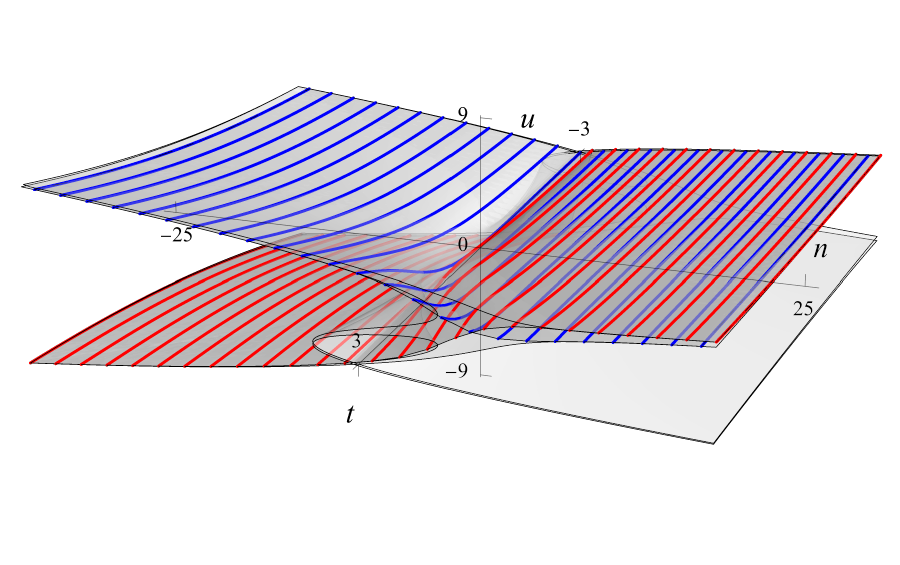}\vspace{-35pt}}
\caption{Globally regular solution of dP$_\text{I}$--P$_\text{IV}$ at   
$-c=b=-1/3$, as a function of $n$ and $t$.}
\label{fig:dP1-P4-global-3D} 
\end{figure}

Numerical experiments indicate that, apparently, the indicated conditions are also sufficient. Figs.~\ref{fig:dP1-P4-global} and \ref{fig:dP1-P4-global-3D} show regular solutions corresponding to several admissible values of the parameter $b=-c$. As we can see, their general behavior differs little from the case with zero termination presented in the previous section; at the same time, these are more complicated, transcendental solutions that cannot be expressed in terms of special functions.

In conclusion of this section, it should be noted that the requirement of regularity for all $t$ is too restrictive. The solutions described above do not exhaust all quasi-stationary solutions of dP$_\text{I}$ at a fixed $t$. For example, the solutions shown in Fig.~\ref{fig:dP1-bctn} are quasi-stationary, although none of the parameters $\beta_n$ vanishes for them. This is explained by the fact that the evolution of these solutions in $t$ leads to poles, which are all located at $t<0$. The family of solutions with zero termination and the initial condition (\ref{u-1b}) has the same property. 

\begin{figure}[h!]
\centerline{\includegraphics[width=\textwidth]{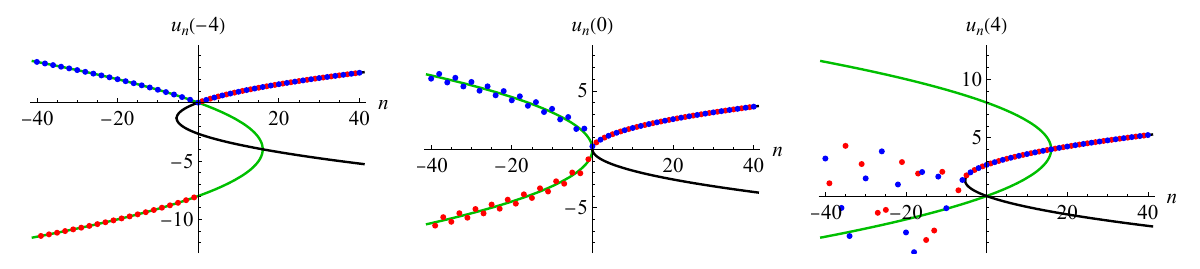}}
\caption{Solution with the asymptotics of type $B$ as $t\to+\infty$.}
\label{fig:dP1-P4-B} 
\end{figure}

Fig.~\ref{fig:dP1-P4-B} shows one more example of a solution, for the parameters $b=c=0$. In it, the functions have the asymptotics of the following types: $u_n\sim(D,B)$ for $n>0$, $u_0\sim(0,B)$, and, for $n<0$, the asymptotics of the types $(A,B)$ and $(C,B)$ alternate. As we have seen, the presence of the type $B$ asymptotics necessarily leads to the appearance of poles, but it turns out that they are all located on the half-line $t>0$. As a result, for $t>0$ a sharp loss of quasi-stationarity is observed when leaving the upper branch of the parabola $C_1$. This regime was studied in the paper by Aptekarev and Novokshenov \cite{Aptekarev_Novokshenov_2024}. However, for $t<0$ there are no poles, and the solution flows over, in a quasi-stationary way, from the parabola $C_1$ to the parabola $C_2$. 

\section{Asymptotics in \texorpdfstring{$n$}{n} for a fixed \texorpdfstring{$t$}{t}}\label{s:n-series}

In this section we construct the series which, in a certain sense, are dual to the series of Section \ref{s:t-series}: now $t$ plays the role of a fixed parameter, and we consider the expansions in the variable $\nu=n-b$. In principle, they can be constructed in the inverse powers of $\sqrt{\pm\nu}$ (as in the paper \cite{Lew_Quarles_1983}), but it turns out to be more practical to use those roots which arise in solving the quadratic equations (\ref{curve1}), (\ref{curve2}) defining the quasi-stationary parabolas $C_1$ or $C_2$. This is justified not only for $c=0$, but also in the general case, since the general quasi-stationary curves (\ref{curve2pm}) coincide asymptotically with these parabolas for large $n$.

\subsection{Series for \texorpdfstring{$n\to+\infty$}{n -> +inf}}

Let $\nu=n-b$ be a large positive number. Introduce the variable
\[
 z=\sqrt{3\nu+t^2}
\]
and construct the solution of dP$_\text{I}$ as an alternating sequence of formal series
\begin{equation}\label{unt1}
\begin{aligned}
 u_n=F=\frac{1}{3}z(1+f_1/z+f_2/z^2+\dots) &\quad\text{at even}~n,\\
 u_n=G=\frac{1}{3}z(1+g_1/z+g_2/z^2+\dots) &\quad\text{at odd}~n,
\end{aligned} 
\end{equation}
whose coefficients are functions of $c$ and $t$ (the parameter $b$ enters only in $z$). The alternation is necessary in order to take into account the difference in the formulas caused by the term $(-1)^nc$. The positive value of $z$ corresponds to the solution on the upper branches of the quasi-stationary curves $C^\pm_2$, the negative one corresponds to the lower ones. At the points neighboring $n$, the variable $z$ is replaced by
\begin{equation}\label{zpm1}
 z_{\pm1}=\sqrt{3(\nu\pm1)+t^2}=z\sqrt{1\pm3/z^2}
  =z\left(1 \pm\frac{3}{2z^2} -\frac{9}{8z^4} \pm\frac{27}{16z^6}-\cdots\right),
\end{equation}
and the substitution into dP$_\text{I}$ leads to the functional equations
\begin{equation}\label{FGeq1}
\begin{aligned}
 & 3F(z)(G(z_{-1})+F(z)+G(z_1)-2t)+t^2-z^2+3c=0,\\
 & 3G(z)(F(z_{-1})+G(z)+F(z_1)-2t)+t^2-z^2-3c=0.
\end{aligned}
\end{equation}
In order to construct the solution, one has to perform the substitution of the series (\ref{zpm1}) into the series (\ref{unt1}):
\[
 F\left(z_{\pm1}\right)=\frac{1}{3}z_{\pm1}(1+f_1/z_{\pm1}+f_2/z^2_{\pm1}+\dots)
\]
and analogously for $G$. Since the series $z^{-k}_{\pm1}$ starts with $z^{-k}$, the power $z^{-k}$ appears only in a finite number of terms; therefore the operation of substitution is well defined and leads to a series of the same type as (\ref{unt1}). Besides, the coefficient of $z^{1-k}$ is a linear combination of the coefficients of the original series with the numbers not exceeding $k$, for example
\[
 F(z_{-1})= \frac{1}{3}z\left(1+\frac{f_1}{z}+\frac{2f_2-3}{2z^2}+\frac{f_3}{z^3}+\frac{8f_4+12f_2-9}{8z^4}+\dots\right).
\]
 It follows that, after substituting all the series into (\ref{FGeq1}) and collecting the like powers of $z$, there arises a sequence of recurrence relations of the form
\[
 f_k+2g_k=\Phi(f_1,g_1,\dots,f_{k-1},g_{k-1}),\qquad 
 2f_k+g_k=\Psi(f_1,g_1,\dots,f_{k-1},g_{k-1}),
\]
where the right hand sides are quadratic polynomials. As a result, all $f_k$ and $g_k$ are found uniquely, as polynomials in $t$ and $c$. The first few terms of the series $F$ have the form 
\[
 F= \frac{z}{3}\left(1+\frac{t}{z}-\frac{9c}{2z^2}-\frac{9ct}{z^3}
  +\frac{3(1+9c^2-48ct^2)}{8z^4}+\frac{3t(1+36c^2-96ct^2)}{8z^5}+\cdots\right)
\]
(it is not difficult to compute 100 terms on the computer, and even more). We list some properties of the constructed series.

\begin{proposition}\label{th:Fid}
1) The series $F$ and $G$ differ in the sign of $c$:
\begin{equation}\label{FGsym}
 G(c,t,z)= F(-c,t,z).
\end{equation}
2) The series $F$ satisfies the identities
\begin{equation}\label{Fsym}
 F(-c,it,iz)= iF(c,t,z),\qquad F(c,-t,-z)= -F(c,t,z).
\end{equation}
3) The series (\ref{unt1}) satisfy the Volterra lattice $u'_n=u_n(u_{n+1}-u_{n-1})$, which is equivalent, taking into account the formula (\ref{FGsym}) and the relation $\partial z/\partial t=t/z$, to the identity
\begin{equation}\label{Fteq1}
 \frac{\partial F}{\partial t}(c,t,z)+\frac{t}{z}\frac{\partial F}{\partial z}(c,t,z)
  = F(c,t,z)(F(-c,t,z_1)-F(-c,t,z_{-1})),
\end{equation}
where $z_{\pm1}=z\sqrt{1\pm3/z^2}$.
\end{proposition}
\begin{proof}
1) The system (\ref{FGeq1}) is invariant under the permutation of $F$ and $G$ with the simultaneous change of the sign of $c$, that is, under the change $\tilde F(c,t,z)=G(-c,t,z)$, $\tilde G(c,t,z)=F(-c,t,z)$. This change leaves the leading coefficients of the series (\ref{unt1}) invariant, and, since the subsequent coefficients are determined uniquely, this means that $\tilde F(c,t,z)=F(c,t,z)$, as required. 

2) The proof is analogous: the system (\ref{FGeq1}) is invariant under the change $\tilde F(c,t,z)=-iF(-c,it,iz)$ (accompanied by the same change for $G$, taking into account the previous property). Here, the leading coefficient of the series $F$ does not change and, since the subsequent coefficients are determined uniquely, $\tilde F(c,t,z)=F(c,t,z)$, as required. The second identity is an obvious consequence of the first one.

3) Let $u_n(t)$ be an arbitrary solution of the dP$_\text{I}$ equation
\[
 u_n(u_{n-1}+u_n+u_{n+1})=2tu_n+\nu-(-1)^nc.
\]
Consider for it the residual under the substitution into the Volterra lattice, that is, define the quantities 
\begin{equation}\label{rn}
 r_n = u'_n-u_n(u_{n+1}-u_{n-1}).
\end{equation}
Let us differentiate dP$_\text{I}$ with respect to $t$, using this equality, and eliminate the variables $u_{n\pm2}$ arisen in this way by means of the dP$_\text{I}$ equation at the points $n\pm1$. It is not difficult to check that this leads to the following equation for the residual:
\[
 (u_{n-1}+u_n+u_{n+1})r_n+ u_n(r_{n-1}+r_n+r_{n+1})=2tr_n.
\]
Now let the variables $u_n$ be given by the series (\ref{unt1}). It is not difficult to see that then the quantity (\ref{rn}) is also a series in the inverse powers of $z$, starting with a certain power, which we denote $k$:
\[
 r_n = \frac{a_k(t)}{z^k}+\frac{a_{k+1}(t)}{z^{k+1}}+\dotso.
\]
Taking into account that $z_{\pm1}=z+O(z^{-1})$, the leading terms of the expansions of $u_{n-1},u_n,u_{n+1}$ coincide and are equal to $z/3$. In exactly the same way, the leading terms of the expansions of $r_{n-1},r_n,r_{n+1}$ coincide and are equal to $a_k/z^k$. But, then, from the equation written out above, the leading term gives $a_k=0$, which means that the whole series equals zero.
\end{proof}

The use of the properties 1) and 2) helps to compute the coefficients of the series more efficiently. In particular, at $c=t=0$ we have $F(0,0,iz)= iF(0,0,z)$, which means that, in this case, the expansion goes, in fact, in $z^{-4}\propto\nu^{-2}$ (not counting the common factor $z$) and the only nonzero coefficients are $f_{4k}$. 

Property 3) means that the evolution in $t$ governed by the Volterra lattice or, equivalently, by the P$_\text{IV}$ equation, preserves the asymptotics given by the series (\ref{unt1}), that is, does not spoil the quasi-stationarity property. Note that this property is not obvious, since in constructing the series we used only the dP$_\text{I}$ equation itself, without using the P$_\text{IV}$ equation or the lattice (\ref{VL}) at all.

\subsection{Series for \texorpdfstring{$n\to-\infty$}{n -> -inf}}

Let now $\nu=n-b$ be a large negative number. We use the same notation $z,F,G$ as in the previous subsection, but define them in a different way, which should not lead to a misunderstanding. Since the solution of the equation (\ref{curve2}) has the form $u=t\pm\sqrt{t^2-\nu}$, set
\[
 z=\sqrt{t^2-\nu},
\]
and construct the solution of dP$_\text{I}$ in the form of the series
\begin{equation}\label{unt2}
\begin{aligned}
 u_n=t+F=t+z(1+f_1/z+f_2/z^2+\dots) &\quad\text{at even}~n,\\
 u_n=t-G=t-z(1+g_1/z+g_2/z^2+\dots) &\quad\text{at odd}~n.
\end{aligned} 
\end{equation}
In contrast with the formulas (\ref{unt1}), here we do not include the term $t$ into the coefficients $f_1$, $g_1$, and also change the sign in front of the series $G$, since the points tend to the upper and lower branches of the curves $C^\pm_2$ alternately (more precisely, if $z>0$, then even $n$ correspond to the upper branch of $C^+_2$, and odd ones correspond to the lower branch of $C^-_2$; if $z<0$, then even $n$ correspond to the lower branch of $C^+_2$, and odd ones correspond to the upper branch of $C^-_2$). When shifting $n$ by $\pm1$, the variable $z$ is replaced by
\[
 z_{\pm1}=\sqrt{t^2-\nu\mp1}=z\sqrt{1\mp1/z^2}
  =z\left(1\mp\frac{1}{2z^2}-\frac{1}{8z^4}\mp\frac{1}{16z^6}-\cdots\right),
\]
and the substitution into dP$_\text{I}$ leads to the system of functional equations
\begin{equation}\label{FGeqm}
\begin{aligned}
 & (t+F(z))(t-G(z_{-1})+F(z)-G(z_1))+z^2-t^2+c=0,\\
 & (t-G(z))(t+F(z_{-1})-G(z)+F(z_1))+z^2-t^2-c=0.
\end{aligned}
\end{equation}
As before, substituting here the series for $z_{\pm1}$ and collecting the like powers of $z$, we obtain a sequence of recurrence equations for the coefficients, even slightly simpler than for the system (\ref{FGeq1}), since it turns out to be solved with respect to $f_k,g_k$ at once:
\[
 f_k=\Phi(f_1,g_1,\dots,f_{k-1},g_{k-1}),\qquad 
 g_k=\Psi(f_1,g_1,\dots,f_{k-1},g_{k-1}),
\]
where the right hand sides are quadratic polynomials. Thus, all $f_k$ and $g_k$ are found uniquely as polynomials in $t$ and $c$. The first few terms of the series $F$ have the form 
\begin{multline*}
 F= z\left(1-\frac{c}{2z^2}-\frac{ct}{z^3}+\frac{1+3c^2}{8z^4}-\frac{(1+4c^2)t}{8z^5}
 +\frac{5c+3c^3-4(1+6c^2)t^2}{16z^6} 
 +\frac{c(13+12c^2)t}{8z^7} \right.\\ \left.
 -\frac{24+138c^2+45c^4+160(c+c^3)t^2}{128z^8}
 +\frac{(77+488c^2+144c^4)t-32c(11+16c^2)t^3}{128z^9} +\cdots\right).
\end{multline*}
The properties of the constructed series are listed in the following proposition. Its proof is quite similar to the proof of Proposition \ref{th:Fid} and may be omitted.

\begin{proposition}\label{th:Fm-id}
1) The series $F$ and $G$ are related by
\begin{equation}\label{FGm-sym}
 G(c,t,z)= -F(-c,t,-z).
\end{equation}
2) The series $F$ satisfies the identities
\begin{equation}\label{Fm-sym}
 F(-c,it,iz)= iF(c,t,z),\qquad F(c,-t,-z)= -F(c,t,z).
\end{equation}
3) The series (\ref{unt2}) satisfy the Volterra lattice $u'_n=u_n(u_{n+1}-u_{n-1})$, which is equivalent to the identity
\begin{equation}\label{Fm-t}
 1+\frac{\partial F}{\partial t}(c,t,z)+\frac{t}{z}\frac{\partial F}{\partial z}(c,t,z)=
  (t+F(c,t,z))(F(-c,t,-z_1)-F(-c,t,-z_{-1})),
\end{equation}
where $z_{\pm1}=z\sqrt{1\mp1/z^2}$.
\end{proposition}

Besides the theoretical value for the justification of the asymptotics of the quasi-stationary solutions, the series (\ref{unt1}) and (\ref{unt2}) turn out to be useful for the numerical computation as well. Like the series of Section \ref{s:t-series}, they diverge but, nevertheless, for large $n$ they give very good approximations to the solution if one considers the partial sum terminated at the term of the minimal absolute value. This makes it possible to make a very precise choice of the initial data for the Cauchy problem at large $n$. Let us return to the experiment described at the end of Section \ref{s:dP1}, where the solutions close to the parabola $C_1$ on a small interval were constructed by choosing the initial data right on it (see Fig.~\ref{fig:dP1-plus}). Let us repeat the same, choosing now the initial data by means of the series (\ref{unt1}). It turns out that taking into account even the first few corrections noticeably lengthens the interval of quasi-stationarity, and if the initial data are computed with the accuracy of about a hundred decimal digits, then this makes it possible to reproduce the solutions shown in Fig.~\ref{fig:dP1-P4-B}. In the same way, the use of the series (\ref{unt2}) makes it possible to obtain the globally regular solutions shown in Fig.~\ref{fig:dP1-P4-global}. In practice, this turns out to be an even more effective method than solving boundary value problems for P$_\text{IV}$.

\section{Series in the transition region \texorpdfstring{$\nu\sim t^2$}{nu ~ t2}}\label{s:self}

\begin{figure}[b!]
\centerline{\includegraphics[width=0.98\textwidth]{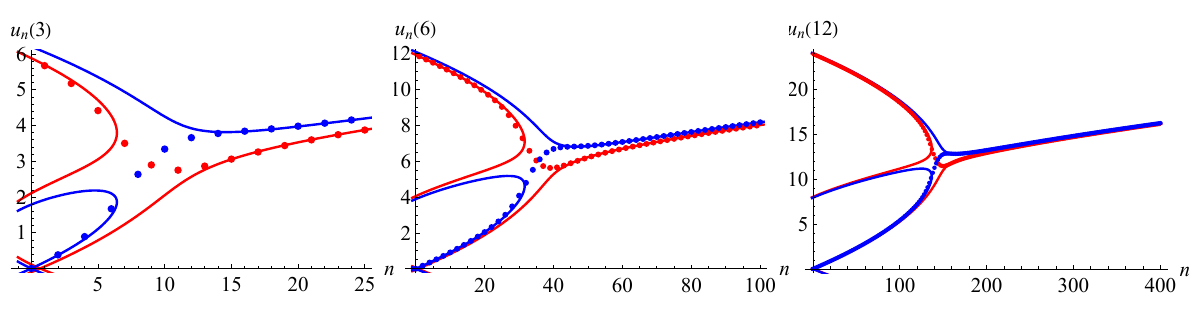}}
\caption{Self-similarity of solutions of dP$_\text{I}$.}
\label{fig:dP1-self} 
\end{figure}

Considering the solutions in the quadrant $\nu>0$, $t>0$, one can notice that the profile of the solution almost does not change under the scaling $(\nu,t,u)\mapsto(\alpha^2\nu,\alpha t,\alpha u)$, where $\alpha>0$ is an arbitrary number, see Fig.~\ref{fig:dP1-self} (as a concrete example, it shows the solution with zero termination at $b=1/3$, that is, this is the same solution as in Fig.~\ref{fig:dP1-zero-plus}, bottom, for other values of $t$. However, the further considerations are applicable also to solutions without zero termination). For large $t$, after passing to the variables (\ref{uxnu}), the profile of the solution tends to the parabolas (\ref{ux}), which means that one can consider it asymptotically self-similar. Here, the relative width of the transition zone between the two parabolas (that is, near the point $\nu=t^2$) tends to 0, but its absolute width in the original variables $n$ grows. The study of this region in the framework of the Riemann--Hilbert approach was carried out in the paper \cite{Bleher_Its_2003}, for the case $c=0$. We will use the matching method \cite{Il'in}. To describe the transition zone, we introduce the stretched variable
\begin{equation}\label{x}
 x=t^{4/3}(\nu/t^2-1),
\end{equation}
and construct the solution in the form of the series
\begin{equation}\label{unt-self}
\begin{aligned}
 u_n=P=t(1+t^{-2/3}p_1(x)+t^{-4/3}p_2(x)+\dots) &\quad\text{at even}~n,\\
 u_n=Q=t(1+t^{-2/3}q_1(x)+t^{-4/3}q_2(x)+\dots) &\quad\text{at odd}~n.
\end{aligned} 
\end{equation}
Let us show how to arrive at the change (\ref{x}) and what conditions on the coefficients $p_1$ and $q_1$ arise in this way. After the change 
\begin{equation}\label{chant}
 z=t\sqrt{1-\rho},\qquad\nu=t^2\rho,
\end{equation} 
the series (\ref{unt2}) for even $n$ is rewritten in the form of the following series in the inverse powers of $t$
\begin{equation}\label{perf}
 u_n=t(1+\sqrt{1-\rho}) +\frac{P_1(\sqrt{1-\rho})}{(1-\rho)t}
  +\frac{P_2(\sqrt{1-\rho})}{(\sqrt{1-\rho})^5t^3}
  +\frac{P_3(\sqrt{1-\rho})}{(1-\rho)^4t^5}+\cdots,
\end{equation}
where
\begin{gather*}
 P_1=-c\left(1+\frac{\sqrt{1-\rho}}{2}\right),\quad 
 P_2=-\frac{1+6c^2}{4}-\frac{(1+4c^2)\sqrt{1-\rho}}{8}+\frac{(1+3c^2)(1-\rho)}{8},\\ 
 P_3= -\frac{c(11+16c^2)}{4}-\frac{5c(1+c^2)\sqrt{1-\rho}}{4}+\frac{c(13+12c^2)(1-\rho)}{8}
      +\frac{c(5+3c^2)(\sqrt{1-\rho})^3}{16},\dotso.
\end{gather*}
At the point $\rho=1$, the coefficients of this series have increasing singularities. Therefore, according to the matching method \cite{Il'in}, near this point one has to pass from the variable $\rho$ to the stretched variable 
\[
 x=(\rho-1)t^{\kappa},
\]
requiring that, after such a stretching, the terms of the series (\ref{perf}) be of one and the same order in $t$ in the main term (except for the first term $t$). This gives the condition $t^{3\kappa /2-2}=1$, from which we find that one should choose $\kappa=4/3$; this leads exactly to the formula (\ref{x}). As a result of this stretching, the series (\ref{perf}) takes the form of the first of the series (\ref{unt-self}). Quite similarly, the same change in the second of the series (\ref{unt2}) leads to the second series of (\ref{unt-self}). Here, the coefficients of both transformed series are, in turn, formal series in the inverse powers of $\sqrt{-x}$. In particular, the coefficients $p_1(x)$ and $q_1(x)$ have the form
\begin{align}
\label{asp}
 p_1(x)&=\sqrt{-x}+\frac{c}{x}-\frac{1+6c^2}{4(\sqrt{-x})^5}
  +\sum_{j>2}\frac{p_{j,1}(c)}{(\sqrt{-x})^{3j-1}},\\
\label{asq}
 q_1(x)&=-\sqrt{-x}-\frac{c}{x}+\frac{1+6c^2}{4(\sqrt{-x})^5}
  +\sum_{j>2}\frac{q_{j,1}(c)}{(\sqrt{-x})^{3j-1}},
\end{align}
where $p_{j,1}(c)$ and $q_{j,1}(c)$ are uniquely determined polynomials in the parameter $c$. According to the matching method, the asymptotics of the solution $u_n$, valid in a neighborhood of the point $\rho=1$, should be constructed in the form of the series (\ref{unt-self}) with the coefficients $p_j(x)$ and $q_j(x)$ smooth for all $x$; moreover, the complete asymptotic expansions of $p_1(x)$ and $q_1(x)$ as $x\to -\infty$ must be determined by the formulas (\ref{asp}), (\ref{asq}). The sign of the square root $\sqrt{-x}$ in these formulas is chosen based on which branch of the quasi-stationary curve the variables with even indices approach as $t\to+\infty$: the upper branch of $C^+_2$ corresponds to the plus sign, and the lower branch (as shown on Fig.~\ref{fig:dP1-self}) corresponds to the minus sign. For the sake of definiteness, in what follows, we shall consider the positive value of the square root.

Several coefficients of the series (\ref{unt-self}) are given in the following proposition. We restrict ourselves to the first three, since the expressions for the subsequent ones are rather cumbersome. 

\begin{proposition}\label{th:self}
A common solution of the dP$_\text{I}$ equation and the Volterra lattice in the form of formal series of the form (\ref{unt-self}) exists if and only if the coefficient $p_1=p$ is an arbitrary solution of the ODE 
\begin{equation}\label{peq}
 p''=\frac{1}{2}p^3+\frac{x}{2}p+c.
\end{equation}
All the other coefficients are expressed uniquely as polynomials in $p,p'$ and $x$; in particular,
\begin{gather*}
 q_1=-p,\qquad q_2=p_2= \frac{1}{4}(p^2+x),\\
 q_3=-p_3= \frac{3}{20}(p')^3 -\frac{9p^4+18xp^2+72cp-23x^2}{240}p'+\frac{3p^3-13xp+18c}{240}. 
\end{gather*}
\end{proposition}

\begin{proof}
The recurrence scheme for the coefficients is, in this case, more involved compared with the series of Section \ref{s:n-series}. Recall that there it was sufficient to use only the dP$_\text{I}$ equation for their computation, and the fact that the result satisfies the Volterra lattice was obtained automatically. Now one has to use both equations at once, but it is more convenient to do this by substituting the series not into them, but into the P$_\text{IV}$ equation (\ref{P4}) and into the relations (\ref{BT-gen}) defining the B\"acklund transformations. This is equivalent, since dP$_\text{I}$ and the Volterra lattice are derived from this set of equations and vice versa. In the substitutions one should take into account, according to the formula (\ref{x}), that 
\[
 \nu=t^2(xt^{-4/3}+1),\quad dx/dt= -2x/(3t)-2t^{1/3},\quad 
 x_{\pm1}=t^{4/3}((\nu\pm1)/t^2-1)=x\pm t^{-2/3}.
\] 
Substituting into P$_\text{IV}$ the series $P$, for even $n$, and collecting the coefficients of the powers of $t$, we obtain, in the leading order, the equation (\ref{peq}) for $p_1$, and the subsequent orders yield a sequence of relations of the form
\begin{equation}\label{pxx}
 4p''_j-2(3p^2_1+x)p_j= \text{~a differential polynomial in~} p_{j-1},\dots,p_1,x.
\end{equation}
Next, rewrite the relations (\ref{BT-gen}) in the form
\[
 u_n= \frac{u'_{n-1}-u^2_{n-1}+2tu_{n-1}+\beta_{n-1}}{2u_{n-1}},\qquad
 u_n= \frac{-u'_{n+1}-u^2_{n+1}+2tu_{n+1}+\beta_{n+1}}{2u_{n+1}}.
\]
Taking $n$ odd, replace in the first equation $u_n$ by the series $Q(t,x)$, and $u_{n-1}$ by $P(t,x-t^{-2/3})$. This will give, after expanding the right hand side in a series in $t^{-2/3}$, explicit expressions for $q_j$ in the form of differential polynomials in $p_{j-1},\dots,p_1,x$. If, however, one considers the difference of these two equations, then again relations only for $p_j$ arise and, keeping track of the highest derivatives, we obtain that they have the form
\[
 4p'''_j-2(3p^2_1+x)p'_j-2(6p_1p'_1-2j+3)p_j
  = \text{~a differential polynomial in~} p_{j-1},\dots,p_1,x.
\]
Subtracting from this the differentiated relation (\ref{pxx}), we obtain an explicit expression for $p_j$. As a result, all the coefficients $p_j,q_j$ are expressed through $p,p'$ polynomially, as claimed. Substituting the found $p_j$ into (\ref{pxx}) gives an identity in view of (\ref{peq}), and no other relations for $p$ arise.
\end{proof}

\begin{figure}[t!]
\centerline{\includegraphics[width=0.8\textwidth]{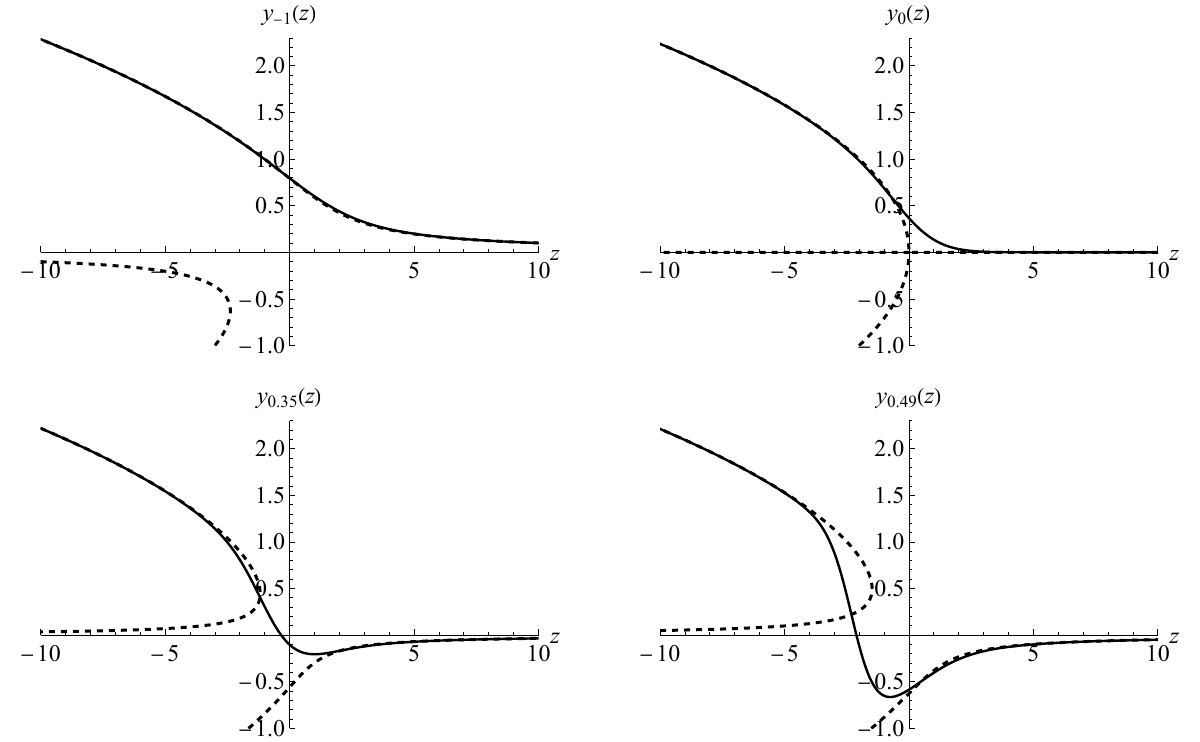}}
\caption{Hastings--McLeod type solutions of the P$_\text{II}$ equation (\ref{P2}), for the parameter values\\ $c=-1$, $c=0$ (the Hastings--McLeod solution itself), $c=0.35$ and $c=0.49$. The dashed line corresponds to the quasi-stationary curve $2y^3+zy+c=0$.}
\label{fig:P2-HM-gen} 
\end{figure}

The change $p(x)=2^{2/3}y(z)$, $z=2^{-1/3}x$ reduces (\ref{peq}) to the P$_\text{II}$ equation
\begin{equation}\label{P2}
 y''=2y^3+zy+c.
\end{equation}
The requirement that the solution of the equation (\ref{peq}) we need have the asymptotics (\ref{asp}) as $x\to -\infty$ leads to the conclusion that, provided the solution $p_1(x)$ is smooth on the whole axis $x$, it coincides, for $c<1/2$, with a generalization of the known special Hastings--McLeod solution \cite{Hastings_McLeod_1980}. This generalization is described in Theorem 1.1 of the paper \cite{Claeys_Kuijlaars_Vanlessen_2008}. Plots of the solution for different values of the parameter are shown in Fig.~\ref{fig:P2-HM-gen}. For $c\leqslant0$ this solution is positive and monotone \cite{Clerc_Davila_Kowalczyk_Smyrnelis_Vidal-Henriquez_2017}.

\begin{remark} 
More precisely, Theorem 1.1 of \cite{Claeys_Kuijlaars_Vanlessen_2008} states that, for any real $c<1/2$, there exists a unique solution of this equation smooth for all real $x$, which has the asymptotics $p_1\sim\sqrt{-x}$ as $x\to-\infty$, and $\lim_{x\to+\infty}x p_1(x)=-2c$. The fact that, as $x\to -\infty$, this solution actually possesses the complete asymptotic expansion (\ref{asp}) follows from Lemma 1.1 of the paper \cite{Il'in_Suleimanov_2004} and the general Kuznetsov theorem \cite{Kuznetsov_1972}, \cite{Malgrange_1972} (see, for more details, the statement in Remark 4 of the paper \cite{Adler_Suleimanov_2026} concerning the asymptotics as $x\to -\infty$ of this solution for $c=0$). Similarly, using Lemma 1.1 of \cite{Il'in_Suleimanov_2004} and the general Kuznetsov theorem, one proves that, as $x\to\infty$, the complete asymptotic expansion of this generalization of the Hastings--McLeod solution of the equation (\ref{peq}), in the case $c\neq0$, $c<1/2$, has the form   
\begin{equation}\label{aspp}
 p_1=-\frac{2c}{x}+\sum_{j=2}^{\infty}\frac{a_j}{x^j}. 
\end{equation}
For $c=0$, the Hastings--McLeod solution tends to zero exponentially as $x\to\infty$.
\end{remark}

Besides, the following Proposition is valid.

\begin{proposition}\label{th:bol}
Any solution $y(z)$ of the equation {\normalfont(\ref{P2})}, regular for all $z$, which has the positive square-root asymptotics $y\sim\sqrt{-z/2}$ as $z\to-\infty$, necessarily has the asymptotics
\[
 y(z)\sim-\frac{c}{z}\quad(c\ne0),\qquad y(z)=o(z^{-1})\quad(c=0),\qquad z\to+\infty.
\]
\end{proposition}

\begin{proof}
Let a real solution of (\ref{P2}) be regular for all sufficiently large positive $z$. We show that it is bounded as $z\to+\infty$. Here $c$ can be any real number.

Choose $M>1$ so that the inequality $M^3>|c|$ holds. For $z\ge0$ and $y\ge M$ we have
\begin{equation}\label{cubic}
 y''=2y^3+zy+c\ge y^3.
\end{equation}
If at some point $z_0$ simultaneously $y(z_0)\geqslant M$ and $y'(z_0)\geqslant 0$, then, under the further continuation, the solution $y$ increases and remains not less than $M$. Indeed, as long as $y\geqslant M$, we have
\[
 y'(z)\geqslant y'(z_0)+M^3(z-z_0)>0 \qquad(z>z_0)
\]
and, after one more integration,
\[
 y(z)\geqslant y(z_0)+y'(z_0)(z-z_0)+\frac{M^3}{2}(z-z_0)^2>M.
\]
It remains to justify that the condition $y\geqslant M$ will not be violated. It holds in a certain right neighborhood of $z_0$. This is obvious if $y(z_0)>M$; if $y(z_0)=M$, then it follows from $y'(z_0)>0$, and, if $y'(z_0)=0$, from $y''(z_0)>0$.

Assume now that the solution later drops below $M$. Then there exists the first return point $z_1>z_0$ to the level $M$, with $y(z)>M$ for $z_0<z<z_1$. But the estimate above, applied on $[z_0,z_1]$, gives
\[
 y(z_1)\geqslant M+\frac{M^3}{2}(z_1-z_0)^2>M,
\]
which contradicts $y(z_1)=M$.

Multiplying further (\ref{cubic}) by $y'\geqslant 0$ and integrating, we obtain
\[
 y'(z)^2\geqslant \frac12y(z)^4-C.
\]
Under the assumption of regularity on the whole right half-axis, the inequality $y''\geqslant M^3$ first ensures an unbounded growth of $y$, and then the last estimate gives $y'\ge y^2/2$ for sufficiently large $y$. Denote by $Y=y(z_1)$ such a fixed positive level that, for $z\geqslant z_1$, the estimate $y'\geqslant y^2/2$ holds. Integration gives
\[
 \frac{1}{Y}-\frac{1}{y(z)} =\int_{z_1}^{z}\frac{y'(t)}{y(t)^2}\,dt \geqslant\frac{z-z_1}{2}.
\]
Hence,
\[
 0<\frac1{y(z)} \leqslant \frac{1}{Y}-\frac{z-z_1}{2}.
\]
At $z=z_1+2/Y$, the right hand side is equal to zero which gives a contradiction. Thus, a positive solution with the estimate $y'\geqslant y^2/2$ cannot exist on the infinite right interval.

Consequently, such a set of values $y,y'$ at the point $z=z_0$ is incompatible with regularity on the whole right half-axis. Hence, on any interval where $y>M$, necessarily $y'<0$. But it is impossible to stay in this region infinitely long: the inequality $y''\ge M^3$ would force the derivative to become positive. Hence, the solution leaves it and can no longer cross the level $M$ upwards.

Applying the same arguments to $-y$, we obtain
\[
 |y(z)|\leqslant M
\]
for all sufficiently large $z$. Consequently, any solution of (\ref{P2}) globally smooth for all $z$ is necessarily bounded as $z\to+\infty$. From this boundedness there already follows the exact leading asymptotics
\[
\lim_{z\to+\infty}zy(z)=-c.
\]
Indeed, fix an arbitrary $\varepsilon>0$ and introduce the barrier functions  
\[
 U(z)=\frac{-c+\varepsilon}{z},\qquad L(z)=\frac{-c-\varepsilon}{z}.
\]
For the upper barrier function
\[
 2U^3+zU+c-U'' =\varepsilon+O(z^{-3})\geqslant\frac{\varepsilon}{2}
\]
for sufficiently large $z$. Set $w=y-U$. Then
\[
 w''=\bigl(z+2(y^2+yU+U^2)\bigr)w+\bigl(2U^3+zU+c-U''\bigr).
\]
Since
\[
y^2+yU+U^2\ge0,
\]
in each region where $w>0$ there holds
\[
 w''\geqslant\frac{\varepsilon}{2}.
\]
A bounded function $w$ cannot remain positive up to $+\infty$ under such an inequality (integrating it twice, we obtain
\[
 w'(z)\geqslant w'(a)+\frac{\varepsilon}{2}(z-a),\qquad
 w(z)\geqslant w(a)+w'(a)(z-a)+\frac{\varepsilon}{4}(z-a)^2;
\]
the right hand side of the second inequality tends to $+\infty$, even if $w'(a)<0$, which contradicts the boundedness of $w$). It cannot have a positive component with two finite zero endpoints either: this would contradict the convexity. Consequently, for sufficiently large $z$
\[
y(z)\leqslant U(z).
\]
Analogously, considering $L-y$, we obtain
\[
L(z)\leqslant y(z).
\]
Thus,
\[
-c-\varepsilon\le zy(z)\leqslant-c+\varepsilon,
\]
which proves the required limit. In particular,
\[
 y(z)\sim-\frac{c}{z}\quad(\text{for~~}c\ne0),\qquad y(z)=o(z^{-1})\quad(\text{for~~}c=0).\qedhere
\]
\end{proof}

For the remaining values of $c$, it turns out that the solutions $y(z)$ with the positive square-root asymptotics as $z\to -\infty$ cannot be regular on the whole axis $z$.

\begin{proposition}  
For $c\geqslant1/2$, all solutions of the ODE {\normalfont(\ref{P2})} with the asymptotics $y\sim\sqrt{-z/2}$ as $z\to-\infty$ necessarily have poles. 
\end{proposition} 

\begin{proof}
We begin with the threshold case $c=1/2$. Set 
\[
 v(z)=y'(z)-y(z)^2-\frac z2.
\]
A direct differentiation using the equation (\ref{P2}) gives
\[
 v'=y''-2yy'-\frac12=-2yv.
\]
Therefore, for a fixed point $z_0$ of the original interval of regularity
\begin{equation}\label{v}
 v(z)=v(z_0)\exp\left(-2\int\limits_{z_0}^{z}y(s)\,ds\right).
\end{equation}
The validity of the asymptotic expansion (\ref{asp}) implies that
\[
 -2\int\limits_{z_0}^{z}y(s)\,ds
 =\frac{2\sqrt2}{3}(-z)^{3/2}(1+o(1)),\qquad z\to-\infty.
\]
So, if $v(z_0)\ne0$, then $|v(z)|$ grows faster than any power of $|z|$. We show that this is impossible. From the asymptotics (\ref{asp}) and the equation (\ref{P2}) we obtain that, as $z\to -\infty$, the estimates
\[
 y(z)=O((-z)^{1/2}),\qquad y''(z)=O((-z)^{3/2})
\]
hold. For a sufficiently large negative $z$, the whole interval $[z,z+1]$ belongs to the original interval of regularity. The exact Taylor formula
\[
 y(z+1)-y(z)=y'(z)+\int_0^1(1-s)y''(z+s)\,ds
\]
gives $y'(z)=O((-z)^{3/2})$. Thus, $v(z)=O((-z)^{3/2})$ as $z\to-\infty$, which contradicts the equality (\ref{v}) for $v(z_0)\ne0$. Hence, $v\equiv0$ and $y(z)$ is a solution of the Riccati equation
\[ 
 y'=y(z)^2+z/2,
\]
which, by the substitution $y=-u'/u$, reduces to the linear equation
\begin{equation}\label{linear}
 u''+\frac{z}{2}u=0.
\end{equation}
Its general real solution is expressed in terms of the classical Airy functions
\[
 u(z)=\gamma_1\operatorname{Ai}(-az)+\gamma_2\operatorname{Bi}(-az),\qquad a=2^{-1/3},
\]
where $\gamma_1,\gamma_2\in\mathbb R$ and $(\gamma_1,\gamma_2)\ne(0,0)$.

As $x\to+\infty$, the asymptotics
\[
\frac{\operatorname{Ai}'(x)}{\operatorname{Ai}(x)}\sim-\sqrt{x},\qquad
\frac{\operatorname{Bi}'(x)}{\operatorname{Bi}(x)}\sim\sqrt{x},\qquad
\frac{\operatorname{Ai}(x)}{\operatorname{Bi}(x)}\to0,\qquad
\frac{\operatorname{Ai}'(x)}{\operatorname{Bi}'(x)}\to0
\] 
hold. Hence $\gamma_2\ne0$. (Then, as $z\to -\infty$  
\[
 y(z)= -\frac{u'(z)}{u(z)} \sim a\sqrt{-az} =\sqrt{-z/2},
\]
while, for $\gamma_2=0$, the sign of the leading term of this asymptotics would be opposite.) Any nonzero real solution of the linear equation (\ref{linear}) has infinitely many zeros as $z\to+\infty$: this follows, for example, from the Sturm comparison theorem with the equation $w''+w=0$ for $z\ge2$.

All zeros of $u$ are simple, since the simultaneous equalities $u(z_*)=u'(z_*)=0$ would imply $u\equiv0$. Therefore, to each zero there corresponds a genuine pole of the first order of $y(z)$. 

Now consider the case of the inequality $c>1/2$. Set
\[
 F(z)=y'(z)-y(z)^2-\frac z2.
\]
The Painlev\'e equation (\ref{P2}) implies the exact equality
\[
 F'+2yF=c-\frac12.
\]
Refining the given asymptotics as $z\to -\infty$, including the derivative, gives
\[
 y(z)=\sqrt{-z/2}+\frac{c}{2z}+O(|z|^{-5/2}), \qquad
 F(z)=\frac{c-\tfrac12}{\sqrt{-2z}}+O(|z|^{-2}).
\]
Therefore, for $c>1/2$ and sufficiently large negative $z$, we have the inequality $F(z)>0$.

This inequality is preserved under the motion to the right, as long as the solution remains regular: at any point where $F=0$, the inequality
\[
 F'=c-\frac12>0,
\]
holds, so that the passage from positive values to negative ones is impossible. Consequently,
\[
 y'>y^2+\frac z2.
\]
Assume now that the solution is regular on the whole real axis. For $z\geq0$
\[
y'(z)>\frac z2,\qquad y(z)>y(0)+\frac{z^2}{4}.
\]
Hence, there exists $z_1>0$ such that $y(z_1)>0$. Further, $y'>y^2$, whence
\[
 \left(\frac1y\right)'<-1,\qquad \frac1{y(z)}<\frac1{y(z_1)}-(z-z_1).
\]
The right hand side of the last inequality vanishes in finite time, which contradicts the global regularity of the positive $y$.
\end{proof}

We now show that, for $c<1/2$, after the rearrangement of the solution of dP$_\text{I}$ in the Painlev\'e II layer, the variables $u_n$ with even indices enter an asymptotic regime described, in the leading order, by the quasi-stationary curve $C^+_2$, while the solutions $u_n(t)$ with odd indices enter an asymptotic regime which is given, in the leading order, by the curve $C^-_2$. For this purpose, we show that the asymptotics (\ref{aspp}) (its leading term $-2c/x$ is found from the quasi-stationary part $(p^3+xp)/2+c=0$ of the second Painlev\'e equation (\ref{peq})) for
\begin{equation}\label{obls}
 \gamma_1t^{-4/3+\delta} < \rho-1< \gamma_2t^{-4/3+\delta}
\end{equation} 
(where $\gamma_1$, $\gamma_2$ are positive real constants, and $\delta$ is an arbitrary real constant from the interval $(0,1/3)$) is consistent with the behavior, as $t\to\infty$, of the quasi-stationary curve $C^+_2$ defined by the corresponding equation (\ref{curve2pm}). 

Indeed, after the change $u=tU$ and the second of the changes (\ref{chant}), this case of (\ref{curve2pm}) takes the form of the equation
\begin{equation}\label{eq:original}
\begin{gathered}
 P(U-1;t,\rho-1)=3(U-1)^4+4(U-1)^3 +2(\rho-1)(2(U-1)+(U-1)^2)-(\rho-1)^2+\\
 \frac{c}{t^2}\bigl(8+12(U-1)+2(\rho-1) +6(U-1)^2\bigr)-\frac{c^2}{t^4}=0. 
\end{gathered}
\end{equation}
The left hand side of this equation, at $U=1/2$ and $\rho-1>\gamma_0t^{-4/3+\delta}$ (where $0<\gamma_0<\gamma_1$), for sufficiently large values of $t$ takes the form of a quantity which is certainly negative   
 \[
 P(-1/2;t,\rho-1)=-\frac{5}{16}-\frac{3}{2}(\rho-1)-(\rho-1)^2 
 +\frac{c}{2t^2}(7+4(\rho-1))-\frac{c^2}{t^4}<0.
 \]
At the same time, for the values $U\geqslant 1/2$, the derivative of this polynomial with respect to $U$ 
\[
 P_U(U-1,t,\rho-1)=4U\left(3(U-1)^2+\rho-1+\frac{3c}{t^2}\right),
\]
satisfies the inequality
\begin{equation}\label{proizv}
 P_U(U-1,t,\rho-1)\geqslant(\rho-1)>0
\end{equation}
(since, by virtue of the inequality $\rho-1>\gamma_0(t^{-4/3+\delta})$, the quantity $t^{-2}(\rho-1)^{-1}$ tends to zero uniformly, and
\[ 
 \rho-1+\frac{3c}{t^2}\geqslant \frac{\rho-1}{2}
\] 
for sufficiently large $t$). Therefore, from the fact that, as $U\to\infty$, the value of the polynomial $P(U-1,t,\rho-1)$ is determined, in the leading order, by the term $3U^4$, it follows that, on the ray $U\in (1/2,\infty)$, there exists a unique solution $U(t,\rho-1)$ of the equation (\ref{eq:original}). For sufficiently large $T$, at $t\geqslant T$ and $\rho-1>\gamma_0t^{-4/3+\delta}$, it is obviously a smooth function of its arguments.

Set 
\[
U_0=1-\frac{2c}{t^2(\rho-1)}. 
\]
A direct substitution shows that, on the interval (\ref{obls}),
\begin{gather*}
 P(U_0-1;t,\rho-1)= -(\rho-1)^2-\frac{32c^3}{t^6(\rho-1)^3}+\frac{24c^3}{t^6(\rho-1)^2}+\\
 (\rho-1)^2(2ch-16c^2B-c^2h^2+48c^4B^2),
\end{gather*}
where, as $t\to\infty$,
\[ 
 h=\frac{1}{t^2(\rho-1)}=O(t^{-2/3-\delta}),\qquad B=\frac{1}{t^4(\rho-1)^3}=O(t^{-3\delta}). 
\]
Therefore 
\[
 P(U_0-1;t,\rho-1)=O\left((\rho-1)^2+\frac{1}{t^6(\rho-1)^3}\right)
 =O\left(t^{-2}(t^{-3\delta}+t^{-2/3+2\delta})\right).
\]
Then, from the Lagrange mean value theorem and from the left hand side of the inequality (\ref{proizv}), it follows
\[
 |U-U_0|\leqslant\frac{|P(U_0-1,t,\rho-1)|}{\rho-1}
  =O\left(\frac{t^{-3\delta}+t^{-2/3+2\delta}}{t^2(\rho-1)}\right).
\]
Consequently, in the strip (\ref{obls}), the unique solution of the equation (\ref{curve2pm}) in the case $C^+_2$ indeed has the asymptotics 
\[
 u=t-\frac{2c}{t(\rho-1)} +O\!\left(\frac{t^{-3\delta}+t^{-2/3+2\delta}}{t(\rho-1)}\right)
  =\left(t-\frac{2ct^{1/3}}{x}\right)\left(1+O(t^{-3\delta}+t^{-2/3+2\delta})\right),
\]
uniform in this strip. In this strip we have
\[
 \gamma_1t^{\delta} < x< \gamma_2t^{\delta},
\]
hence, as $t\to \infty$ and $x\to \infty$, the validity of the asymptotics (\ref{aspp}) for $p_1(x)$ implies that also
\[
 t+t^{1/3}p_1(x)= \left(t-\frac{2ct^{1/3}}{x}\right)\left(1+O(t^{-\delta})\right). 
\]
Consequently, in the strip (\ref{obls}), the sum $t+t^{1/3}p_1(x)$ is matched precisely with the quasi-stationary curve $C^+_2$. In a similar way, one shows that the sum $t+t^{1/3}q_1(x)$ is matched, in this strip, with the quasi-stationary curve $C^-_2$.

\section{Concluding remarks}\label{s:conclusion}

We have examined the problem of finding solutions to the dP$_\text{I}$ and P$_\text{IV}$ equations that exhibit a specific quasi-stationary asymptotic behavior. In the special case of zero termination, this problem has a long and rich history rooted in the theory of orthogonal polynomials, and a lot of beautiful results. Relaxing the termination condition and extending the solutions to negative values of $n$ makes the problem considerably more complex. We have constructed formal asymptotic expansions for these quasi-stationary solutions with respect to $t$ and $n$ and verified their consistency. Based on the analysis of these expansions, we derived necessary conditions for the existence of global quasi-stationary solutions; our numerical experiments suggest that such solutions do indeed exist. A rigorous proof of this fact remains an open problem. Using the method of matching asymptotic expansions, we obtained a uniform self-similar asymptotic for the transition zone between different quasi-stationary regimes; the result is expressed in terms of special Hastings--McLeod type solutions to the P$_\text{II}$ equation. We also examined several families of exact solutions that exhibit a combination of quasi-stationary and pole behavior. A detailed study and classification of such solutions is also an open problem. Overall, it must be acknowledged that there are currently more open questions than answers, and we do not claim that our results are exhaustive. Nevertheless, we believe that the consistency of the quasi-stationary dynamics of two Painlev\'e-type equations underlying our approach is a conceptually important observation that may also prove useful in a wide range of other models.

\subsection*{Acknowledgments}

The work of V.E.~Adler was carried out within the state assignment FFWR-2024-0012 of the Landau Institute for Theoretical Physics of the Russian Academy of Sciences. The work of B.I.~Suleimanov was carried out within the state assignment FMRS-2025-0012 of the Institute of Mathematics with Computing Centre of the Ufa Federal Research Centre of the Russian Academy of Sciences.



\begin{thebibliography}{99}

\bibitem{Shohat_1939} J. Shohat. A differential equation for orthogonal polynomials. {\em Duke Math. J. \bf 5}(2), \href{https://doi.org/10.1215/S0012-7094-39-00534-X}{401--417} (1939).

\bibitem{Freud_1976} G. Freud. On the coefficients in the recursion formulae of orthogonal polynomials. {\em Proc. Roy. Irish Acad. Sect. A \bf 76}, \href{https://www.jstor.org/stable/20489026}{1--6} (1976).

\bibitem{Bonan_Nevai_1984} S. Bonan, P. Nevai. Orthogonal polynomials and their derivatives, I. {\em J. Approx. Theory \bf 40}(2), \href{https://doi.org/10.1016/0021-9045(84)90023-6}{134--147} (1984).

\bibitem{Lew_Quarles_1983} J.S. Lew, D.A. Quarles, Jr. Nonnegative solutions of a nonlinear recurrence. {\em J. Approx. Theory \bf 38}(4), \href{https://doi.org/10.1016/0021-9045(83)90154-5}{357--379} (1983).

\bibitem{Clarkson_Jordaan_Kelil_2015} P.A. Clarkson, K. Jordaan, A. Kelil. A generalized Freud weight. {\em Studies in Appl. Math. \bf 136}, \href{https://doi.org/10.1111/sapm.12105}{288--320} (2016).

\bibitem{Clarkson_Jordaan_2018} P.A. Clarkson, K. Jordaan. Properties of generalized Freud polynomials. {\em J. Approx. Theory \bf 225}, \href{https://doi.org/10.1016/j.jat.2017.10.001}{148--175} (2018).

\bibitem{Alsulami_Nevai_Szabados_Assche_2015} S.M. Alsulami, P. Nevai, J. Szabados, W. Van Assche. A family of nonlinear difference equations: existence, uniqueness, and asymptotic behavior of positive solutions. {\em J. Approx. Theory \bf 193}, \href{https://doi.org/10.1016/j.jat.2014.04.012}{39--55} (2015).

\bibitem{Brezin_Kazakov_1990} E. Brezin, V. Kazakov. Exactly solvable field theories of closed strings. {\em Phys. Lett. B \bf 236}(2), \href{https://doi.org/10.1016/0370-2693(90)90818-Q}{144--150} (1990).

\bibitem{Its_Kitaev_Fokas_1990} A.R. Its, A.V. Kitaev, A.S. Fokas. The isomonodromy approach in the theory of two-dimensional quantum gravitation. {\em Russ. Math. Surveys \bf 45}(6), \href{https://doi.org/10.1070/RM1990v045n06ABEH002699}{155--157} (1990).

\bibitem{Fokas_Its_Kitaev_1991} A.S. Fokas, A.R. Its, A.V. Kitaev. Discrete Painlev\'e equations and their appearance in quantum gravity. {\em Commun. Math. Phys. \bf 142},  \href{https://doi.org/10.1007/BF02102066}{313--344} (1991).

\bibitem{Fokas_Its_Kitaev_1992} A.S. Fokas, A.R. Its, A.V. Kitaev. The isomonodromy approach to matrix models in 2D quantum gravity. {\em Commun. Math. Phys. \bf 147}, \href{https://doi.org/10.1007/BF02096594}{395--430} (1992).

\bibitem{Magnus_1995} A.P. Magnus. Painlev\'e-type differential equations for the recurrence coefficients of semi-classical orthogonal polynomials. {\em J. Comput. Appl. Math. \bf 57}(1--2), \href{https://doi.org/10.1016/0377-0427(93)E0247-J}{215--237} (1995).

\bibitem{Cresswell_Joshi_1999} C. Cresswell, N. Joshi. The discrete first, second and thirty-fourth Painlev\'e hierarchies. {\em J. Phys. A: Math. Gen. \bf 32}(4), \href{https://doi.org/10.1088/0305-4470/32/4/009}{655--669} (1999).

\bibitem{Lukashevich_1967} N.A. Lukashevich. Theory of the fourth Painlev\'e equation. {\em Differential equations \bf 3}(5), \href{https://www.mathnet.ru/rus/de/v3/i5/p771}{771--780} (1967) (in Russian).

\bibitem{Gromak_Laine_Shimomura_2002} V.I. Gromak, I. Laine, S. Shimomura. Painlev\'e differential equations in the complex plane. Berlin: Walter de Gruyter, 2002.

\bibitem{Veselov_Shabat_1993} A.P. Veselov, A.B. Shabat. Dressing chains and the spectral theory of the Schr\"odinger operators. {\em Funct. Anal. Appl. \bf 27}(2), \href{https://doi.org/10.1007/BF01085979}{81--96} (1993).

\bibitem{Adler_1994} V.E. Adler. Nonlinear chains and Painlev\'e equations. {\em Physica D \bf 73}(4), \href{https://doi.org/10.1016/0167-2789(94)90104-X}{335--351} (1994).

\bibitem{Bassom_Clarkson_Hicks_1995} A.P. Bassom, P.A. Clarkson, A.C. Hicks. B\"acklund transformations and solution hierarchies for the fourth Painlev\'e equation. {\em Stud. Appl. Math. \bf 95}, \href{https://doi.org/10.1002/sapm19959511}{1--71} (1995). 

\bibitem{Gromak_1978} V.I. Gromak. Single-parameter families of solutions of Painlev\'e equations. {\em Differential Equations \bf 14}(12), \href{https://www.mathnet.ru/eng/de/v14/i12/p2131}{2131--2135} (1978) (in Russian).

\bibitem{Bassom_Clarkson_Hicks_McLeod_1992} A.P. Bassom, P.A. Clarkson, A.C. Hicks, J.B. McLeod. Integral equations and exact solutions for the fourth Painlev\'e equation. {\em Proc. Roy. Soc. London Ser. A \bf 437}(1899), \href{https://www.jstor.org/stable/52049}{1--24} (1992). 

\bibitem{Bassom_Clarkson_Hicks_1993} A.P. Bassom, P.A. Clarkson, A.C. Hicks. Numerical studies of the fourth Painlev\'e equation. {\em IMA J. Appl. Math. \bf 50}(2), \href{http://dx.doi.org/10.1093/imamat/50.2.167}{167--193} (1993). 

\bibitem{Nevai_1984} P. Nevai. Asymptotics for orthogonal polynomials associated with $\exp(-x^4)$. {\em SIAM J. Math. Analysis \bf 15}(6), \href{https://doi.org/10.1137/0515092}{1177--1187} (1984).

\bibitem{Bleher_Its_1999} P. Bleher, A. Its. Semiclassical asymptotics of orthogonal polynomials, Riemann--Hilbert problem, and universality in the matrix model. {\em Annals of Math. \bf 150}(1), \href{https://doi.org/10.2307/121101}{185--266} (1999).

\bibitem{Bleher_Its_2003} P.M. Bleher, A.R. Its. Double scaling limit in the random matrix model: the Riemann--Hilbert approach. {\em Comm. Pure Appl. Math. \bf 56}(4), \href{https://doi.org/10.1002/cpa.10065}{433--516} (2003).

\bibitem{Ercolani_Lega_Tippings_2022} N. Ercolani, J. Lega, B. Tippings. Dynamics of nonpolar solutions to the discrete Painlev\'e I equation. {\em SIAM J. Appl. Dyn. Sys. \bf 21}(2), \href{https://doi.org/10.1137/21M1445156}{1322--1351} (2022).

\bibitem{Suleimanov_1994} B.I. Suleimanov. Onset of nondissipative shock waves and the ``nonperturbative'' quantum theory of gravitation. {\em J. Exp. Theor. Phys. \bf 78}(5),  \href{http://jetp.ras.ru/cgi-bin/e/index/e/78/5/p583?a=list}{583--587} (1994).

\bibitem{Cosgrove_2000} C.M. Cosgrove. Higher-order Painlev\'e equations in the polynomial class I. Bureau symbol $P2$. {\em Studies in Appl. Math. \bf 104}(1), \href{https://doi.org/10.1111/1467-9590.00130}{1--65} (2000).

\bibitem{Adler_Suleimanov_2026} V.E. Adler, B.I. Suleimanov. Self-similar asymptotics in the decay problem for the Volterra lattice with zero boundary condition. arXiv:\href{http://arxiv.org/abs/2606.13137}{2606.13137} (2026).

\bibitem{Schiff_Twiton_2024} J. Schiff, M. Twiton. Classification of real solutions of the fourth Painlev\'e equation. {\em Mathematics \bf 12}, \href{https://doi.org/10.3390/math12030463}{463} (2024).

\bibitem{Clarkson_Loureiro_Assche_2016} P.A. Clarkson, A.F. Loureiro,  W. van Assche. Unique positive solution for an alternative discrete Painlev\'e I equation. {\em J. Diff. Eqs. Appl. \bf 22}(5), \href{http://dx.doi.org/10.1080/10236198.2015.1127917}{656--675} (2016).

\bibitem{Clarkson_Dzhamay_Hone_Mitchell_2025} P.A. Clarkson, A.V. Dzhamay, A.N.W. Hone, B. Mitchell. Special solutions of a discrete Painlev\'e equation for quantum minimal surfaces. {\em Theor. Math. Phys. \bf 224}(2), \href{https://doi.org/10.1134/S0040577925080045}{1359--1397} (2025).

\bibitem{Aptekarev_Novokshenov_2024} A.I. Aptekarev, V.Yu. Novokshenov. Asymptotics of solutions of the discrete Painlev\'e I equation. {\em Math Notes \bf 116}(6), \href{https://doi.org/10.1134/S0001434624110270}{1170--1182} (2024).

\bibitem{Mathematica_2021} Wolfram Research, Inc. Mathematica, Version 13.0. Champaign, IL (2021).

\bibitem{Il'in} A.M. Il’in. Matching of asymptotic expansions of solutions of boundary value problems. Nauka, Moscow, 1989; Amer. Math. Soc., RI, Providence, 1992.

\bibitem{Hastings_McLeod_1980} S.P. Hastings, J.B. McLeod. A boundary value problem associated with the second Painlev\'e transcendent and the Korteweg--de Vries equation. {\em Arch. Rational Mech. Anal. \bf 73}(1), \href{https://doi.org/10.1007/BF00283254}{31--51} (1980).

\bibitem{Claeys_Kuijlaars_Vanlessen_2008} T. Claeys, A.B.J. Kuijlaars, M. Vanlessen. Multi-critical unitary random matrix ensembles and the general Painlev\'e II equation. {\em Annals of Math. \bf 168}, \href{https://doi.org/10.4007/annals.2008.168.601}{601--641} (2008).

\bibitem{Clerc_Davila_Kowalczyk_Smyrnelis_Vidal-Henriquez_2017} M.G. Clerc, J.D. D\'{a}vila, M. Kowalczyk, P. Smyrnelis, E. Vidal-Henriquez. Theory of light-matter interaction in nematic liquid crystals and the second Painlev\'e equation. {\em Calculus of Variations and Partial Differential Equations \bf 56}, \href{https://doi.org/10.1007/s00526-017-1187-8} {93} (2017).

\bibitem{Il'in_Suleimanov_2004} A.M. Il'in, B.I. Suleimanov. Birth of step-like contrast structures connected with a cusp catastrophe. {\em Sbornik: Math. \bf 195}(11--12), \href{https://doi.org/10.1070/SM2004v195n12ABEH000863}{1727--1746} (2004). 

\bibitem{Kuznetsov_1972} A.N. Kuznetsov. Differentiable solutions to degenerate systems of ordinary equations. {\em Funct. Anal. Appl. \bf 6}(2), \href{https://doi.org/10.1007/BF01077515}{119--127} (1972).

\bibitem{Malgrange_1972} B. Malgrange. Sur les points singuliers des \'equations diff\'erentielles. {\em S\'eminaire Goulaouic-Schwartz \bf XX}, \href{https://www.numdam.org/item/SEDP_1971-1972____A20_0.pdf}{1--13} (1972).

\end{thebibliography}
\end{document}